\documentclass{aa}  
\usepackage{epsfig}
\usepackage{graphicx}
\usepackage{xcolor}
\usepackage{txfonts}
\usepackage{amssymb}
\usepackage{amsfonts}
\usepackage{natbib}
\usepackage{hyperref}
\hypersetup{
  colorlinks=true, 
  urlcolor=blue, 
  linkcolor=blue, 
  citecolor=blue 
}

\bibpunct{(}{)}{;}{a}{}{,}

\begin{document}

\title{Confrontation between modelled solar integrated observables and direct observations\\ I. Radial velocities and convective blueshift\thanks{Table \ref{lambda} are only available in electronic form at the CDS via anonymous ftp to cdsarc.u-strasbg.fr (130.79.128.5) or via http://cdsweb.u-strasbg.fr/cgi-bin/qcat?J/A+A/.} }

    \titlerunning{Confrontation between modelled solar integrated radial velocities and direct observation}

\author{N. Meunier\inst{1},  A.-M. Lagrange\inst{2,1},  X. Dumusque\inst{3}, S. Sulis\inst{4}
}
\authorrunning{N. Meunier et al.}

\institute{
Univ. Grenoble Alpes, CNRS, IPAG, F-38000 Grenoble, France \email{nadege.meunier@univ-grenoble-alpes.fr}
\and
LESIA (UMR 8109), Observatoire de Paris, PSL Research University, CNRS, UMPC, Univ. Paris Diderot, 5 Place Jules Janssen, 92195 Meudon, France
\and
Observatoire astronomique de l'Université de Genève, 51 chemin Pegasi, 1290 Versoix, Switzerland
\and 
Universit\'e Aix Marseille, CNRS, CNES, LAM, Marseille, France
 \\
     }

   \date{}


\date{Received ; Accepted}

\abstract{Stellar variability strongly impacts the search for low-mass exoplanets with radial velocity techniques. Two types of planet-free time series can be used to quantify this impact: models and direct solar observations after a subtraction of the Solar System planetary contribution. Making a comparison among these approaches is necessary to improve the models, which can then be used for blind tests across a broad range of conditions. }
{Our objective is therefore to validate the amplitude of the convective blueshift in plages
used in our previous works, particularly in blind tests, with HARPS-N solar data. }
{We applied our model to the structures observed at the time of HARPS-N observations and established a direct comparison between the radial velocity time series. To complete our diagnosis, we also studied the observed radial velocities separately for each diffraction  order derived from the individual cross-correlation functions, as well as our line-by-line radial velocities.}
{We find that our previous model had been underestimating the amplitude of the convective blueshift inhibition by a factor of  about 2. A direct estimation of the convective blueshift in the spectra, which is shown to be correlated with the plage filling factor, allows us to explain the difference with previous estimations obtained with MDI/SOHO Dopplergrams, based on the specific properties of the Ni line used in this mission. In addition, we identified  several instrumental systematics, in particular, the presence of a 2 m/s peak-to-peak signal with a period of about 200 days in radial velocity and bisector. This signal could be due to periodic detector warm-ups, a systematic dependence  of the long-term trend on wavelength that is possibly related to the variability of the continuum over time, and/or an offset in radial velocity after the interruption of several months in October 2017.}
{A large amplitude in the convective blueshift inhibition of (360~m/s, namely twice more than in our previous works) must be used when building synthetic times series for blind tests. The presence of instrumental systematics should also be taken into account when using sophisticated methods based on line properties to mitigate stellar activity when searching for very weak signals.}


\maketitle
%

\section{Introduction}

Stellar variability has a major impact on radial velocity (RV) measurements 
and it is currently the dominant limitation involved in detection studies of low-mass planets around solar-type stars using this technique. A good understanding of these limitations requires  realistic blind tests to be performed by injecting fake planets to evaluate their recovery. This can be done either on the basis of synthetic time series \cite[][]{dumusque16,dumusque17,meunier23}  or on observed solar time series, such as those obtained with the High Accuracy Radial velocity Planet Searcher for the Northern hemisphere \cite[HARPS-N,][]{dumusque15,phillips16,collier19,dumusque21} at the Galileo National Telescope (TNG) or with other instruments such as the High Accuracy Radial velocity Planet Searcher (HARPS) on the European Southern Observatory (ESO) 3.60~m telescope or the EXtreme PREcision Spectrometer (EXPRES) at the 4.3 m Lowell Discovery Telescope and soon the Echelle SPectrograph for Rocky Exoplanets and Stable Spectroscopic Observations (ESPRESSO) at the Very Large Telescope (VLT). This can be done after the subtraction of the contribution from the Solar System  planets, which is  well-known; indeed, these are the only series for which we are sure that they are planet-free. Observed solar RVs also have the advantage to include all physical processes: They have  been used  to understand better solar RVs, focusing, for example, on short-term variability \cite[][]{almoulla23} for making comparison with other instruments \cite[][]{zhao23b} or various activity indicators \cite[][]{sen23}, as well as to test mitigation techniques  \cite[][]{lienhard22,zhao22,debeurs22} or new formalisms \cite[][]{hara23}. However, they are of limited duration (a few years so far) and contain gaps, along with being representative of the Sun only. In addition, despite the fact that such direct observations   basically represent  the ground truth about solar RV variations, they can be impacted by specific effects due to the finite size of the Sun and to the fact that the Earth  is orbiting around the Sun, leading to some spurious effects, such as those  corrected for in \cite{collier19}. However, there could be additional effects present in the data. It is then important to also be able to rely on synthetic time series, allowing any temporal sampling and adaptation to  other spectral types or activity levels. For that purpose, we must compare the models with such  direct solar observations, so that we may validate the prescription used for the different processes and to identify any  process impacting RVs that may be missing from the synthetic time series. 

In this paper, we focus on the convective blueshift inhibition in plages \cite[][]{dravins81,dravins82,dravins86}, which causes part of the rotation modulation as well as variability at longer timescales, noting that the rotation modulation is also due to the contrast of spots and plages. Other processes play an important role, such as granulation and supergranulation \cite[][]{meunier19e,meunier20b} and meridional circulation \cite[][]{meunier20c}, as reviewed   in \cite{meunier21}; however, the convective blueshift inhibition in plages was found to be dominant in the solar case \cite[][]{meunier10a,meunier10}. 

We proposed the model in \cite{meunier10a} with the aim to reconstruct the solar RVs due to the temperature contrast in spots and plages, as well as the convective blueshift inhibition in plages. 
The prescription from this paper, which was at the time validated based on an RV reconstruction \cite[][]{meunier10}, has been derived from MDI/SOHO Dopplergrams \cite[][]{Smdi95} that had also been used to generate solar synthetic time series \cite[][]{borgniet15} as well as synthetic time series for other stars \cite[][]{meunier19}. 
Those were used to characterise the amplitude of the RV jitter due to these processes \cite[][]{meunier19b} and to perform  blind tests \cite[][]{meunier23}. These authors showed that for Earth-mass planets in the habitable zone around such stars,  it is difficult to reach a precision of 10\% on the mass estimation in RV follow-ups of transit detections with common techniques. In addition, there is a large level of false positives when performing a blind search, in part due to some long-term residuals after correction.

 The objective of the present paper is to compare our model and prescriptions with HARPS-N solar data and identify whether effects that are not included in our model have an impact on the observed RVs. The outline of the paper is as follows. We first describe the HARPS-N observations and our models in Sect.~2. The direct comparison between RV time series is presented in Sect.~3, with the objective to evaluate and improve the prescription for the convective blueshift inhibition. In Sect.~4, we analyse the variability of the convective blueshift based on a direct approach. Finally, in Sect.~5 , we focus on the long-term variability, mostly on the 200 day period found in the RV time series, thereby we have been able to study different sources of instrumental systematics. We present our conclusions in Sect.~6.

\section{Observations and models}

\subsection{HARPS-N observations}

We used three years of the publicly available HARPS-N solar RV times-series. We note that these data were downloaded using the dace-query python API \footnote{\texttt{https://dace-query.readthedocs.io/en/latest/dace\_\\query.sun.html}}, which gives us access to an upgraded data reduction compared to what was published in Dumusque et al. 2021 (data reduced with ESPRESSO pipeline version 2.3.5 instead of 2.2.3). In addition to what was published and available on the Data \& Analysis Center for Exoplanets (DACE) website\footnote {\texttt{https://dace.unige.ch/sun/?}} \cite[data reduction with the new ESPRESSO pipeline, which features the most stable wavelength solution and corrections for effects due to the Sun not being point-like and observed from inside the Solar System; see ][]{collier19}, {the Data Reduction Software (DRS)} 2.3.5 provides for HARPS-N improved  flat-field corrections, better quality flag to reject bad spectra, and smaller long-term trend systematics (X. Dumusque, private communication). We note that the RVs are extracted using a classical cross-correlation function approach.

We used 603 days of observations covering about 3 years, from July 2015, 29 to July 2018, 16 (a few observations obtained on 11 May 2016 were rejected because they were obvious outliers). These observations were performed during the descending phase of cycle 24, which has a relatively low amplitude  (Fig.~\ref{serie}). The maximum spot number from SILSO/SIDC was 103 for cycle 24, whereas it was 169 for cycle 23. 
The whole data set corresponds to a total of 30995 spectra, with 2 to 106 spectra (and an average of 51) per day. In the following, time corresponds to barycentric Julian day minus 2450000. The significant gap between days 8069 and 8136 is due to a broken optical fibre injecting sunlight into the HARPS-N calibration unit. 
In addition to the RV time series, we mostly used the S-index, from the original 2021 release, retrieved from 
the DACE\footnote{\texttt{https://dace.unige.ch/sun/}} platform), in which instrumental systematics have been removed \cite[][]{dumusque21}. {The S-index is defined as the flux integrated in the core of the Ca II H \& K lines divided by the continuum, providing information  on the chromospheric emission. We also used (} to a lesser extent) the bisector span (BIS) and the full width at half maximum (FWHM) of the CCF, both from the most recent reduction.  
In addition, we retrieved the S1D spectra and the CCFs for the different {diffraction orders (produced by the grating spectrometer)}, which are described in Sects.~4 and 5 in the context of recomputing the RVs in for subsets of lines.

\subsection{Models}

\subsubsection{2010 simulations, MDI reconstruction, and selections }

The main objective of this paper is to test the validity of the prescription for the inhibition of the convective blueshift in plages used in \cite{meunier10a}. In this previous work, the RVs were computed based on the following procedure, which was then applied to cycle 23. The observed solar structures were localised on a disk, and a spectrum was attributed to each pixel depending on whether it was a spot, a plage or quiet Sun. This allowed us to compute the integrated spectrum corresponding to each contribution \cite[][]{chelli00,galland05}. The spot and plage contrast as well as the contribution of the convective blueshift inhibition in plages were computed separately and then summed up. The prescription for the amplitude of the inhibition of the convective blueshift in plages was 190 m/s, applied perpendicular to the solar surface and globally (i.e. to the whole spectrum).

These reconstructed RV time series were then compared to a more direct estimation of RVs \cite[][]{meunier10}: We reconstructed the solar RVs by integrating the MDI  Dopplergrams \cite[][]{Smdi95}, obtained in the Ni line at 6768~$\AA$. This estimation did not involve any modelling. Because the zero velocity of the spacecraft was not known sufficiently well, we computed the zero based on the quiet Sun average velocity. As a consequence, these reconstructed RVs corresponded to the active region contribution only, but they could be compared to the reconstruction based on spots and plages, leading to an agreement within 30\%. Based on this agreement, the 190~m/s prescription was then used in our subsequent works \cite[e.g.][]{borgniet15, meunier19}.

\subsubsection{Application of the 2010 and 2019 models to cycle 24}

In principle, it would have been interesting to compare the RV from the 2010 model with direct observations. However, such a comparison would have to rely on an activity indicator such as the S-index. A brief description of such an approach is given in Appendix~\ref{cyc23_24}. As a consequence, we instead applied a similar model to cycle 24 spots and plages to perform a direct comparison between daily modelled RVs and the observed HARPS-N RVs. This requires for the spots and plages to be defined in a  manner similar to what was done in our 2010 protocol. The spot catalogue used in 2010 is not available for the entire duration of the HARPS-N data however, we therefore extracted a list of spots from HMI intensity maps. To define plages, we used HMI magnetograms, which have different properties (spatial resolution, noise level, spectral line) than the MDI magnetograms that were used in 2010. 
 The calibration of the thresholds to define spots and plages to ensure an equivalent definition to previous studies is described in Appendix~\ref{app_calib}.

The protocol was then applied to HMI data for each day of the HARPS-N time series. The details are given in Appendix~\ref{app_cyc24}. There is one day for which no HMI data was available, so that the comparison can be performed for 602 days. For each day, we applied the model used in 2010, and in particular the prescription of 190 m/s for the inhibition of the convective blueshift in plages which we wish to test. The RV time series were computed following the analytical approach described in \cite{meunier19}, and we therefore checked that there is a good agreement with the original series used in 2010 \cite{meunier10a}. 
In Sect.~3,  we consider an edge-on Sun, and the impact of the departure from this configuration is studied in Appendix~\ref{append_solar}.

In addition to the 190~m/s prescription, we also tested two velocity dependencies of the convective blueshift inhibition on plage size. The first one was used in \cite{meunier19}, derived from the MDI analysis of \cite{meunier10}. We also reanalysed the data from 2010 and propose a simple power law modelling this dependence:

\begin{equation}
\label{eqpowerlaw}
\Delta V = 148 \times A^{0.0607}
,\end{equation}

where $\Delta V$ is the local convective blueshift inhibition in plages in m/s and A in ppm. These two approaches are  tested in Sect.~3.2. The details are given in Appendix~\ref{app_depa}. We expect that with a weaker contribution of the network compared to plages, the long-term variability will be slightly lower; however,  the rotational modulation will not be significantly affected because the network structures, which are spread over the surface, do not have a strong impact at this scale.

\section{Direct comparison between RVs: Convective blueshift contribution}

\label{sec3}

\subsection{Comparison at short and long timescales based on the 2010 model}

\begin{figure}
\includegraphics{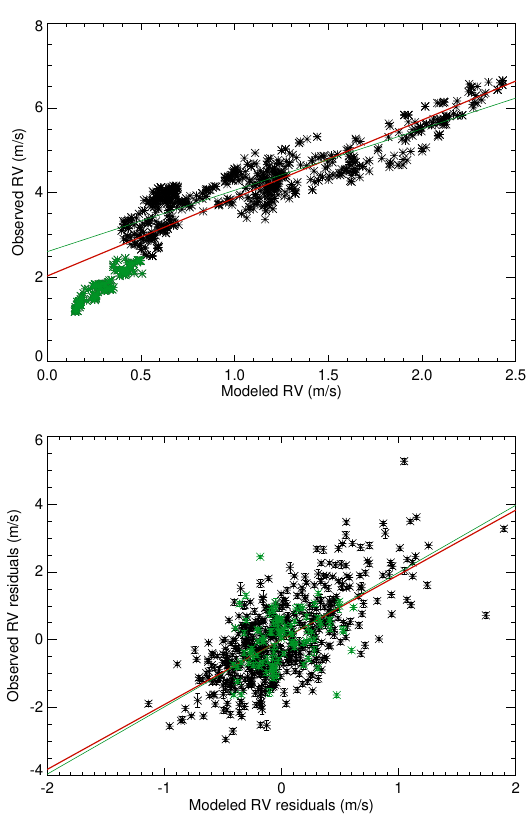}
\caption{Observed binned RV versus modelled binned RV (upper panel) and residuals (lower panel). The binning was done over 28 days. The model uses the 2010 prescription for the amplitude of the convective blueshift inhibition. Red lines are linear fits on all days. Green points correspond to observations performed after the interruption, and the green lines are linear fits without those days.
}
\label{ltct}
\end{figure}

In this section, we compared the model performed with the original prescription of 190~m/s applied to all plages and network structures. We first compared the long-term (LT) and short-term (ST) variability of the two time series and then both scales together. 

We first binned the RVs (model and observed) in 28 days bins to average a large part of the rotational modulation. A linear fit between the observed RVs with respect to the model provides a scaling factor to be applied to the model to fit the observations. We found a slope of 1.85$\pm$0.02, which is significantly larger than 1. The fit is shown in the upper panel of Fig.~\ref{ltct}. This suggests that the convective blueshift was  underestimated in the 2010 model. However, RVs obtained after the facility interruption (shown in green) do not behave as they do in previous observations, according to our model, and with significantly lower RVs, which may bias the fit. Excluding those days, the slope is 1.45$\pm$0.03, still larger than 1. 
The LT variability can be impacted by effects not included in the model, such as a different level in the contribution of the network features (effect which should not affect much the ST slope, see Sect.~3.2, because they are spread over the whole surface), or the presence of meridional circulation \cite[][]{meunier20c}.

To estimate the ST scaling factor, this binned series was subtracted from the unbinned RVs and the residuals analysed in the same way, with a linear fit of the observed residuals versus the model. The fit is illustrated in the lower panel of  Fig.~\ref{ltct}. The slope is 1.91$\pm$0.05 when considering all days. When considering the days before the interruption only, the slope is 1.98$\pm$0.05, namely, larger than the LT slope for the same data selection. The rms of the residuals is 0.87 m/s.

\begin{figure}
\includegraphics{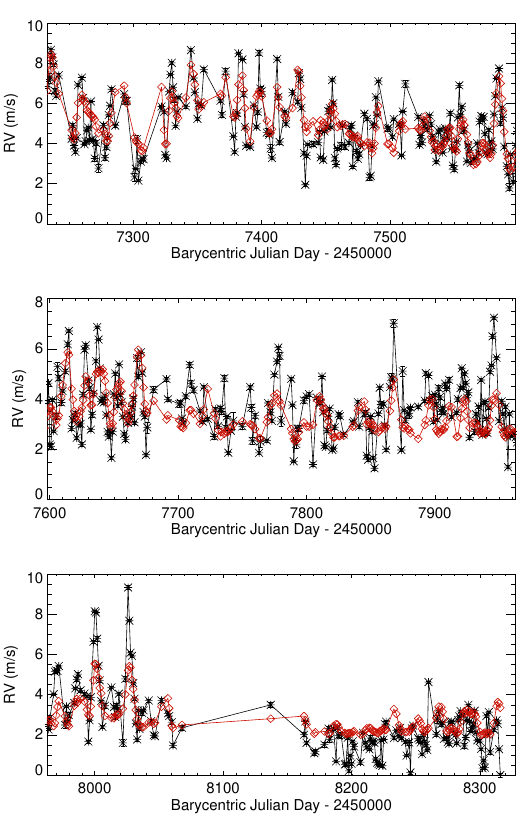}
\caption{Observed (in black) and modelled (in red) RVs versus time. The model is the 2010 version scaled following the three parameter fit in Eq.~\ref{eq3par} and performed over all days (see Sect.~3.1).  
}
\label{rv1zoom}
\end{figure}

\begin{figure}
\includegraphics{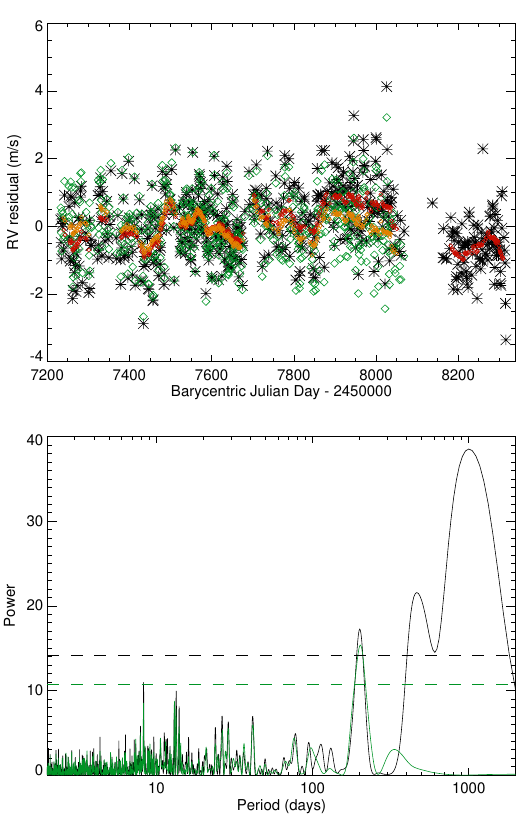}
\caption{Residual RVs versus time (upper panel) and Lomb Scargle periodogram (lower panel) for the 2010 model and scaled following the three parameter fit in Eq.~\ref{eq3par} from Sect.~3.1. The black stars and lines correspond to all days, and the green stars and lines to days before the interruption only. The red (resp. orange) points are the binned (over 28d) residuals for the black (resp. green) points. The horizontal lines in the lower panel correspond to 0.1\% fap level, based on a bootstrap analysis. 
}
\label{rv1}
\end{figure}

Finally, we performed a global fit to take  both LT and ST variability into account, in which we applied a scaling factor  to the contribution due to the inhibition of the convective blueshift, which is the dominant process. 
In principle, an additional correcting factor could also be applied to the spot and plage contrast contribution; however, since this signal is small compared to the additional dispersion due to granulation or supergranulation and since the spot and plage contrast were well constrained in \cite{meunier10a}, we do not expect this contribution to be better constrained here. 
We then adjusted the following model aiming at deriving the scaling factor to be applied to the model that is necessary to properly represent the observations:
\begin{equation}
\label{eq3par}
    RV(t)=RV_{sppl}(t) + \alpha RV_{conv}(t) + \beta t /1000 + k
,\end{equation}
where $RV_{sppl}(t)$ is due to the spot and plage contrast, $RV_{conv}(t)$ is due to the convective blueshift, $\alpha$ is the scaling factor for the convective blueshift, $\beta$ is an additional trend (that is the most simple model, justified by the fact that we have only three years of data, meant to help evaluate whether an additional LT variability is required), and $k$ is a constant. Also, 
$\beta$ provides the corresponding velocity difference over 1000 days (approximately the duration of the observation). The fits were performed based on a least-square minimisation, and the uncertainties computed with a Monte Carlo simulation based on the uncertainties on the observed RVs. 

When applied to all data, we find $\alpha$= 1.774$\pm$0.005 and $\beta$= -0.42$\pm$0.01~m/s. 
Again, when computed over all days, this suggests the need for an additional LT trend over the three years. 
The results are shown in Fig.~\ref{rv1zoom}. There is a general agreement, although a few peaks (in particular between days 7940 and 8030)  are not well fitted.
It is very likely that a simple models for all structures lacks the necessary dispersion. A law that depends on the size of the structure is explored in the next section.

We also performed the same fit with the days before the interruption only. In this case, $\alpha$= 1.944$\pm$0.006, and  $\beta$=1.37$\pm$0.02~m/s. The rms of the residuals is 0.94 m/s. The $\alpha$ value confirms that the prescription was probably underestimated in the 2010 model, and $\beta$ is significantly different from zero. 
This fit with the days before the interruption only leads to a comparison between observed and modelled RVs very similar to Fig.~\ref{rv1zoom}, but the difference between the two after the interruption is slightly reinforced, with a typical offset  of 1.4~m/s. This difference in the behaviour of the RVs after the interruption is also seen when considering a model based on the average absolute magnetic field computed from HMI magnetograms,  as with the Bremen Composite Magnesium II index from the LASP Interactive Solar Irradiance Datacenter (University of Bremen\footnote{\texttt{https://www.iup.uni-bremen.de/gome/solar/\\MgII\_composite.dat}}), but not strongly when using a model based on the S-index.
On the other hand, a few low SNR measurements obtained during the interruption shows a regular decrease during that period, although a temperature warm-up has been performed just before starting again the regular operation, which may also impacts RVs. The origin of the step remains therefore unclear, and we cannot rule out a unusual behaviour of the convective blueshift in the small active regions after the interruption.  

The rms of the residuals, of the order of 1~m/s, is very good given that we expect a large contribution to those residuals coming from processes not taken into account in the model and, in particular, supergranulation. A large amplitude of 0.68~m/s was indeed found by \cite{almoulla23} based on the HARPS-N day-to-day RV dispersion. An analysis of the structure functions by \cite{lakeland23} led to an amplitude of 0.86~m/s. It is also impacted by another effect, as described below. 

The residuals between observation and model (after scaling with  $\alpha$) are shown in the upper panel in Fig.~\ref{rv1}, for both  selections (all days and days before the interruption only), and the periodograms are shown in the lower panel. When considering all days, the residuals show a slope before the interruption, and then the small drop in RV level again. This impacts the periodograms with power for periods above a few hundred days, not seen when considering the days before the interruption only (in green). There is some residual power around Prot/2, hardly significant, which may be due to the fact that some peaks are not perfectly fitted. The dominant feature is a  significant peak at 202 d, corresponding to residuals of almost 2 m/s peak-to-peak. One of our objectives was to study the LT residuals in order to identify which processes were possibly missing, but it is not an easy task because of this peak. \cite{dumusque21} found a similar peak after removing a linear trend to the HARPS-N RVs. We study this peak in more details in Sect.~\ref{sec5}.

We conclude that the 190 m/s prescription underestimates the convective blueshift for the HARPS-N solar RVs by a factor of about 1.9. This means that the MDI reconstruction strongly underestimates the RVs as well in the sense that MDI RVs are not representative of the global spectrum: This apparent discrepancy is discussed in Sect.~3.3. It is also likely that there is an offset in RVs due to the interruption of the facilities during a period of more than two months, which may not be of solar origin.

\subsection{Comparison based on the 2019 model}

We also performed the same analysis with two other modelled time series, the first one with the size-dependent convective blueshift used in \cite{meunier19}, and the second with the power law described in Eq.~\ref{eqpowerlaw} (Sect.~2.2.3). In both cases, the average convective blueshift over all structures (weighted by their size) is equivalent to the 190~m/s prescription  \cite[which we recall was the same for all structures in][]{meunier10a}, in order to only  test  the impact of size-dependent law. 
When considering all points, we found $\alpha$=1.569$\pm$0.005 and $\beta$=-0.77$\pm$0.01 for the first one and $\alpha$=1.725$\pm$0.005 $\beta$=-0.52$\pm$0.01 for the second one. 
When considering only days before the interruption, which is  more reliable given the possible offset, we find $\alpha$=1.719$\pm$0.005 and $\beta$=0.98$\pm$0.02  and $\alpha$=1.892$\pm$0.006 and $\beta$=1.27$\pm$0.02 respectively.

The scaling factors are therefore slightly smaller than when considering a fixed value for the convective blueshift inhibition in plages. Their behaviour is otherwise very similar in terms of a possible  LT trend, which, if considering only the days  before the interruption, is positive. 
Also, the rms of the residuals is similar in all cases, so that it is not possible to discriminate between different size dependencies based on the quality of the fit. However, Eq.~\ref{eqpowerlaw} is probably the most likely solution, as it relies on a  determination of the size dependency of the attenuation of the convective blueshift. In the following, we use a ratio of 1.89 for the correction factor applied to the prescription of 2010. 

In the previous section, we noted that a few peaks were not well fitted. This is still true here, so that a law depending on the size is not sufficient to reproduce the detail of the behaviour of certain regions. A detailed analysis of the properties of active regions present at each time is beyond the scope of the present paper. We checked that the assumptions made on spots (we neglected their convective blueshift to their very small size and low flux, and a constant contrast) could not explain the discrepancies. 
We conclude that the observed departure is probably due to another process, for example a significant dispersion in the convective blueshift inhibition in plages of the same size, based on the average magnetic field or other properties of the region (state of evolution for example), that may lead to a different RV-$\log R'_{HK}$ relationship.

\subsection{Impact on the 2010 MDI reconstruction}
\label{sec33}

\begin{figure}
\includegraphics{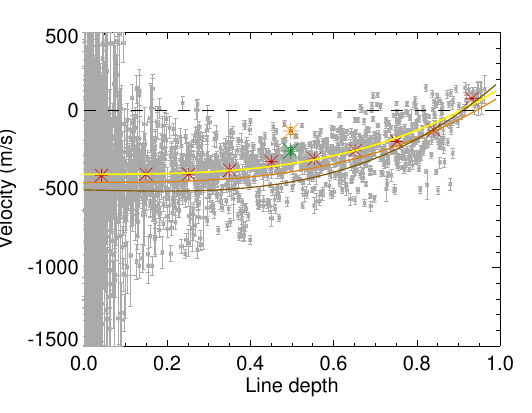}
\caption{Velocities computed with respected to laboratory wavelengths versus line depth. Each point in gray corresponds to a spectral line, averaged over all spectra, after subtraction of a solar gravitational redshift of  {about 636~m/s}. The error bars indicate the 1-$\sigma$ dispersion (and not the error on the average). The orange and green  stars correspond to the 6768 Ni line (MDI/SOHO) and 6173 Fe line (HMI/SDO) respectively. The red stars show the average velocity in each bin in depth and the yellow line is a third degree polynomial fit on those points following the function proposed by \cite{liebing21}. The y-scale has been restricted to {2 km/s}, and a few points are off-scale for very small line depths. The orange line is from \cite{liebing21} for HARPS spectral resolution and after subtraction of a gravitational redshift of {about 636 m/s}, and the brown line is from \cite{reiners16}, based on high resolution spectra.
}
\label{ni}
\end{figure}

The comparison of the 2010 model and the observed HARPS-N RVs shows that the convective blueshift inhibition was likely to have been underestimated by about a factor of $\sim$1.89. This means that the RVs reconstructed with the MDI data were in fact significantly lower than the RVs that would have been derived from all spectral lines such as with HARPS-N. Since we estimated that the MDI RV variability represented 0.88 time the RVs derived from the 2010 model, the MDI variability then represents about 0.47 (0.88 divided by 1.89)  times the new RVs derived from an updated model representative of the HARPS-N DRS RVs.

To understand this factor of 0.48, we compared the behaviour of the Ni line used by MDI (6768 $\AA$) with that of other lines. This line was not included in previous works: \cite{reiners16} considered only Fe I lines, and interpreted the scatter of the convective blueshift for a given line depth to be due to uncertainties on the measurement and blends. The Ni line has however been chosen by the MDI team in part because it was not blended, so any departure from the average behaviour of the lines should not be due to  blends. \cite{liebing21}, based on a larger number of lines (but again without any Ni lines), studied the effect of wavelength on the convective blueshift, and showed that lines at a longer  wavelength exhibited a smaller convective blueshift. In addition, the range covered by the convective blueshift as a function of wavelength was larger for deep lines. This was previously observed by \cite{dravins81} and \cite{hamilton99}. We  confirm this trend with wavelength. The Ni line being on the red side of the optical range, we could expect a smaller convective blueshift using that line than expected from its depth. Other factors may lead to differences, such as the atom or excitation potential.

To quantify this difference, we computed velocities from HARPS-N solar spectra for a large set of lines with respect to their laboratory wavelength. The velocity computation and line selection are described in Appendix~\ref{app_tss}. The average velocity (over all measurements for a given line) versus line depth, $d,$ are represented in Fig.~\ref{ni}, after subtraction of a  gravitational redshift {of about 636 m/s} 
The velocity in bins in $d$ is represented in red, and the yellow curve shows a fit of the function $\gamma \times f(d)$+constant, where $f(d)$ is the function fitted by \cite{liebing21}\footnote{They removed the linear term based on the assumption that the deepest lines form deep enough to correspond to a constant velocity, leading to no gradient for zero absorption depth. The quadratic term was found to be negligible. The final function was found to be sufficient to fit the data. See \cite{liebing21} for more details. } for the HARPS resolution (547.01$d^3$+179.24). We find $\gamma$=0.997, namely, very close to 1, demonstrating a very good agreement with \cite{liebing21}, apart from a small offset in velocity. 
We find that the Ni line, whose measurements are shown in orange, lies indeed above the yellow line in Fig.~\ref{ni} and stands apart: The convective blueshift for the Ni line is -129 m/s instead of -338 m/s for the average line of similar line depth (0.497), corresponding to the velocity of much deeper lines ($d\sim$0.80).

We now wish to compare the Ni line convective blueshift with the convective blueshift equivalent to the HARPS-N DRS.  
An average velocity based on the yellow curve and the lines used in the G2 ESPRESSO mask (used to compute the HARPS-N solar DRS RVs) or with a wavelength dependence \cite[fitted on our data but with a function similar to those in][]{liebing21} was computed. We also tested different line weighting (no weighting, weighting with $d$ and $d^2$). The latter should be more realistic because the DRS weighs the lines according to the RV uncertainty of each spectral line, which at first order, if we consider that all the lines have a similar width, depends on the square of the line depth.
We could not determine what is the best weighting however, because we find little impact of line depth selection on the result, as described in Appendix~\ref{app_lbl}: Both weighting gives global RVs in very good agreement with the DRS RV. A weighting with $d$ gives an equivalent convective blueshift of -267~m/s, and a weighting with $d^2$ gives -222~m/s. They lead to a ratio of 0.48 and 0.58 respectively, to be compared to the 0.48 ratio derived from our comparison between time series.

In addition to this dependence on the weighting, other effects can affect this ratio however. 
First, the RV zero may be uncertain. 
The curve found by \cite{liebing21}, although in excellent agreement concerning the curvature, is for example found at lower velocities by about 30 m/s, while the theoretical convective blueshift found for a small number of lines in \cite{asplund00} in the low $d$ regime ( at high resolution and after scaling to full disk assuming simple projection effects) corresponds to about -330~m/s, which is not very different from what we obtain. In addition, Fig.~\ref{ni} shows several deep lines with an apparent redshift: It is not clear whether this is real or not, but an examination of the selected lines do not show any reversal of the short-term variability (see Appendix~\ref{app_lbl}), suggesting that the global curve could be lower, by up to 100 m/s. 
Second, MDI velocity were computed with a lower spectral resolution, based on four filters with a 75 m$\AA$ bandpass \cite{Smdi95}. However, the RVs derived from these four low resolution filters were corrected \cite[][]{Smdi95} to be in principle representative of high resolution spectra based on MDI simulations (J. Schou, private communication).

We conclude that the particular behaviour of the Ni line used by MDI, which departs from most lines with a similar depth, explains the apparent disagreement between the agreement that was obtained between models and MDI reconstruction in 2010, and the significant difference between the same model and HARPS-N solar data found in the present paper.

On a final note, we also show in Fig.~\ref{ni} (in green) the position of the 6173~$\AA$ line used by HMI. We find that this line is closer to the global curve than the MDI line, with a convective blueshift of -257 m/s. The HMI Dopplergrams were used in \cite{haywood16} and \cite{milbourne19} and \cite{haywood22} to reconstruct solar RVs: They applied a scaling to the resulting time series to establish a correspondence between the RV derived for this specific line and RV computed on a large set of lines. 
\cite{haywood16} found a scaling factor of 1.85$\pm$0.27 and \cite{milbourne19} an averaged scaling factor of 0.93$\pm$0.11, but on the first release of HARPS-N solar data. The scaling factor of \cite{milbourne19} is in better agreement with our convective blueshift, which is close to the value found for both weighting, -222 and -267 m/s.

\subsection{Consequences for planet detectability}
\label{sec34}

We find a larger than expected contribution of the convective blueshift attenuation in plages, by about a factor of 2. Stellar variability was already problematic with the low prescription for this contribution in the estimation of the RV jitter \cite[][]{meunier19b} and in the blind tests performed in \cite{meunier23}: 
A larger convective blueshift leads to worse performance in terms of planet detection capabilities and characterisation. As an example, in a blind test corresponding to a RV follow-up of a transit detection, for a 1~M$_{\rm Earth}$ planet in the middle of the habitable zone of a G2 star, the uncertainty becomes 0.92~M$_{\rm Earth}$ instead of 0.70~M$_{\rm Earth}$ with the previous amplitude. For a blind test to search for a similar planet in RV, the detection rate are decreased by a factor of $\sim$2 (from 3.7~\% to 1.5~\%) and the already high rate of wrong planet detection is slightly increased accordingly.

For comparison, the prescription used by \cite{herrero16} was 300 m/s applied globally to the whole spectrum and derived from hydrodynamical simulations, although they did not simulate plages: this prescription is slightly lower than our new values. In the SOAP simulations of \cite{zhao23}, a prescription of 340 m/s was considered, based on the curve obtained by \cite{liebing21} for a median line depth of 0.7: this value is quite close to our new prescription. 

It is  important to keep in mind that the estimation of the convective blueshift inhibition depends on how the RVs are computed. Any prescription in convective blueshift (and a fortiori in the amplitude of its inhibition in plages) therefore depends on how the RVs are computed (selection of spectral lines and their respective weighting). In addition, reconstructions from Dopplergram may not be representative of the full set of lines.

Finally, if the lower RVs found after the interruption is due to an artefact, then the model based on the active region contribution (after the proper scaling factor) is missing a long-term trend of the order of 1~m/s. \cite{meunier20c} showed that over cycles 22 and 23, meridional circulation could lead to an amplitude of this process of for the Sun seen edge-on between 1 and 2 m/s depending on the cycle. We computed the expected slope for a three year time series during the descending phase of those cycles based on those reconstructed time series and found values between -1.1 and 1.3~m/s: The 1.3~m/s trend is therefore compatible with that range, although it is not well constrained given the degeneracy with the law describing the attenuation of the convective blueshift as a function of size, and the possible systematics (Appendices~\ref{app_ccf} and \ref{app_lbl}).

\section{Evaluating the variability of the convective blueshift inhibition in plages}

\begin{figure}
\includegraphics{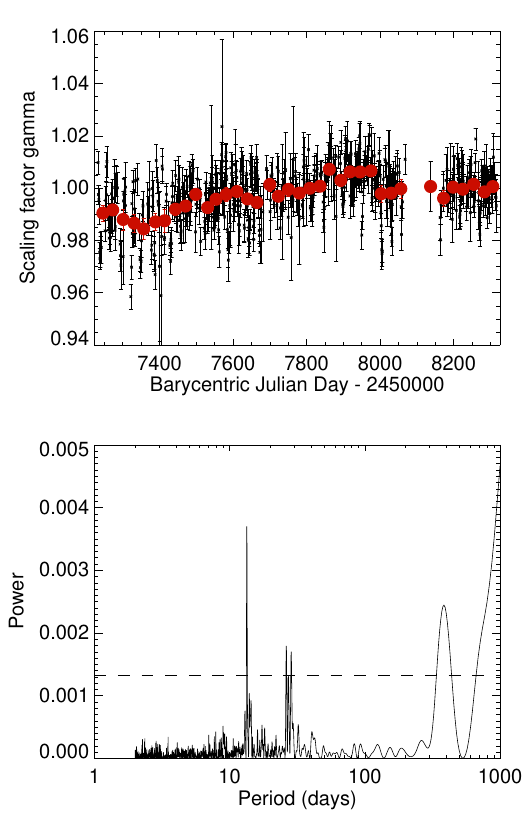}
\caption{Scaling factor, $\gamma,$ versus time and Lomb-Scargle periodogram (all point in black and for observations before the interruption only in green). The red line corresponds to a 28 d binning. The horizontal line on the lower plot is  the 0.1~\% fap level.  }
\label{tss}
\end{figure}

\begin{figure}
\includegraphics{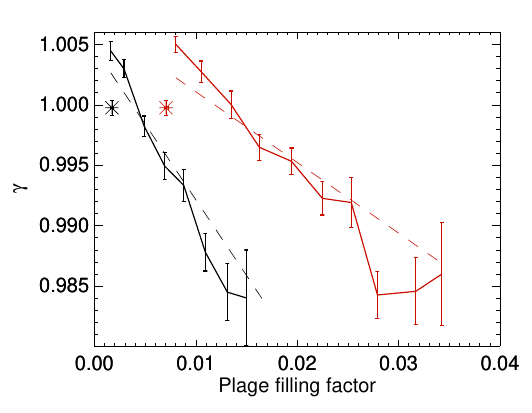}
\caption{Scaling factor, $\gamma,$ versus plage filling factor defined  with a 55~G threshold (black) and a 23~G threshold (red).
}
\label{tss2}
\end{figure}

In Sect.~3.3, we used the velocities with respect to laboratory wavelengths to  characterise the average convective blueshift based on the third signature \cite[][]{gray09} and compare with  the Ni line position. In this section, we study the temporal variation of this third signature, based on a third degree polynomial fit as in \cite{liebing21}, with a single parameter $\gamma$ describing the change in slope (see Sect.~3.3)\footnote{It uses more information that with the slope for line depths above 0.4 used in \cite{meunier17} and \cite{meunier17b}.}. 
Details are given in Sect.~3.3 and Appendix~\ref{app_tss}. 

The $\gamma$ factor, shown as a function of time in Fig.~\ref{tss} (upper panel), exhibits variability above the uncertainties. It first increases until t$\sim$8100, which is expected, since the Sun is progressively less active. 
The average level for $\gamma$ after the interruption is, however, lower, which is not expected: A lower $\gamma$ value should correspond to a more active period (that is more inhibition). There may be an artefact affecting $\gamma$ similar to the RV offset discussed in Sect.~3.

The periodogram (lower panel of Fig.~\ref{tss}), dominated by the long-term trend and with peaks at Prot and Prot/2 that are significant above the 0.1~\% fap level,  so that $\gamma$ is sufficiently precise to exhibit rotational modulation. 
Fig.~\ref{tss2} shows $\gamma$ versus the filling factor (ff) of plages (for two thresholds to define them).  The behaviour of $\gamma$ as a function of the S-index and  |B|$_{\rm disk}$, not shown here, is similar. We observe  
 a significant decrease towards more active configurations. The points corresponding to observations after the interruption are shown apart, exhibiting a different behaviour however.

We discuss now the possibility to use the variability of the  third signature to evaluate the attenuation factor of the convective blueshift in solar plages. We recall that the factor of two-thirds used in \cite{meunier10a} was based on the work of \cite{BS90}, but this factor is otherwise not well documented. The relative variation of this factor as a function of spectral type for a large sample of stars was evaluated in \cite{meunier17b}. We explored two methods, detailed in Appendix~\ref{fact_att}. First, we built a model with two components (quiet and active) to reproduce the dependence of $\gamma$ on the filling factor. The use of the low magnetic field threshold (23~G) provides a lower limit of 0.58$\pm$0.09. On the other hand, a factor of close to 1 was  obtained when exploiting the prescription and average CB obtained in Sect.~\ref{sec3}. 
We conclude that uncertainties remains large, but both estimations point towards a strong attenuation in plages, possibly stronger that was obtained in \cite{BS90}. More work needs to be done to refine those uncertainties and to have better control over the biases.

\section{Analysis of the peak at $\sim$200 d}
\label{sec5}

One objective of this paper is also to identify effects not taken into account in our model. The presence of a strong supergranulation signal is a limitation to the identification of any other short-terms effect, which may affect the use of some proxies. The comparison on short timescales being limited, we therefore do not aim at conducting a complete comparison of activity indicators here, and 
focus on the peak at $\sim$200 d found in the comparison between model and observations and already seen in \cite{dumusque21}. This stands in the way of a  study of the long-term effects involved \cite[e.g. the non-linearity between RVs and S-index found in][due to a combination of projection effects and of the butterfly diagram]{meunier19c} in more details, and therefore need to be understood. The blind tests in \cite{meunier23} also show that even when taking such long-term effects into account, there are still some residuals at long periods strongly affecting the exoplanet detectability and characterisation performance, so that a better understanding of the variability at those long timescales is necessary.

\subsection{Analysis of HARPS-N time series}
\label{sec51}

\begin{figure}
\includegraphics{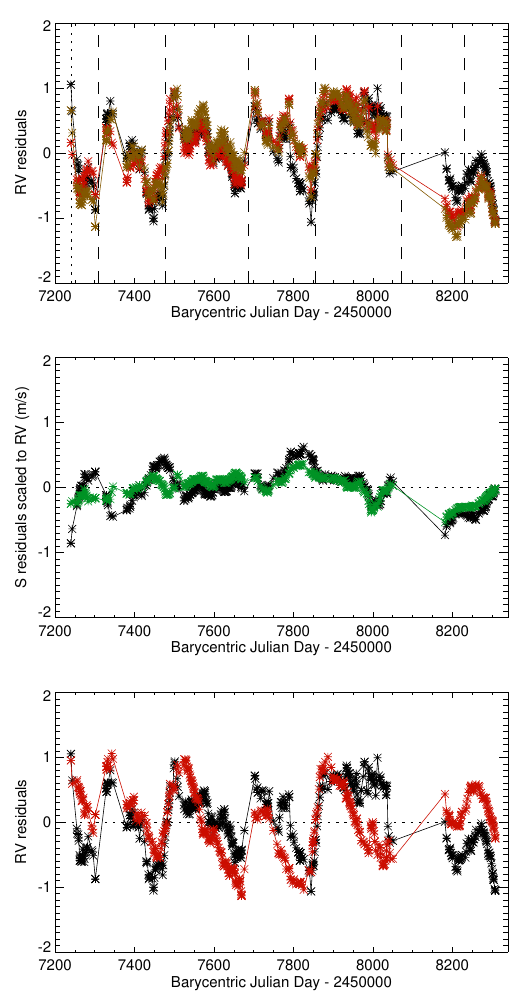}
\caption{RV residuals versus time after subtraction of the models (upper panel), based on the S-index (black), on the ff (red), and on |B|$_{\rm disk}$ (brown). The dashed vertical lines indicate the time of instrument warm-up, and the dotted vertical line the time of power failure which is before day 8100 \cite[][]{dumusque21}. The middle panel shows the S-index residuals scaled to RVs (based on ff in black and on  |B|$_{\rm disk}$  in green), on the same scale than the upper panel. The RV residuals based on the S-index model (in black) are compared to the BIS residuals (in red) in the lower panel. 
}
\label{rvca}
\end{figure}

We first explore the RV variability seen in the HARPS-N observation but not in our model (Fig.~\ref{rv1}) through their relationship with other observables with the S-index, focusing on the variability on a timescale of 200 days. Our objective here is to compare the behaviour of the different observables.  The time series were binned (using a running mean) over one rotation period (28 days), to better visualise the variability at longer time scales. We performed a linear fit of RV versus S-index, computed the resulting model and then subtracted it from the binned RV  to analyse the residuals. 
The superposition of RVs and model, residuals and periodograms are shown in Fig.~\ref{pic200obs}. We also observe a clear modulation, similar to the one observed when comparing with our model in Fig.~\ref{rv1}. The maximum peak in the periodogram is at 202 days, with possibly a secondary peak around 180 d.
The residuals obtained with a RV model based on a linear relationship with the plage filling factor and |B|$_{\rm disk}$ from HMI magnetograms (obtained independently from HARPS-N RVs) provides the same results, with similar amplitude and periods. The peak of the periodogram of the residuals is respectively at 202 and 196 days.

For comparison, we performed the same analysis for the S-index itself, which we model linearly as a function of the plage filling factor and |B|$_{\rm disk}$. 
The residuals are much more irregular than the RV residual,   no significant peak in the periodogram. In addition, if scaled to RVs, it corresponds to a much smaller amplitude, as shown in the middle panel of Fig.~\ref{rvca}. 
A similar analysis on the BIS, which is correlated with the RV signal (correlation of 0.74), also exhibits a strong peak at 200~d, with  residuals (shown in the lower panel of Fig.~\ref{rvca}) similar to the RV residuals (correlation of 0.38), and with a strong amplitude. The rms of the RV residuals after a correction based on a linear relationship between the RV and the BIS time series decreases from 1.97 to 1.32~m/s.
The peak of the periodogram of the residuals is at 182 days. 
The FWHM however is not correlated with RVs nor the S-index (hence a flat model), and is dominated by a peak around 6 months, which has been identified to be due to a variation of the angle between the ecliptic and the solar rotation axis, $B_0$ \cite[][]{collier19}, leading to a variation of the solar $v \sin i$ over time.

We conclude that whatever is the source of this RV signal, it is also visible in the BIS and is therefore affecting line shapes. It is probably not a problem related  to the S-index computation.

\subsection{Comparison with our models and other RV reconstructions}

The same analysis for our cycle 24 model from Sect.~3.1, based on the comparison with the {S-index}, the plage filling factor, and |B|$_{\rm disk}$ shows no such residuals at periods $\sim$200 days present. 
The same is true for our cycle 23 model based on the plage filling factor. A variability is observed at a level of $\pm$0.2 m/s due to the projection effects combined with the butterfly diagram \cite[][]{meunier19c}. The same is true for the RVs reconstructed from MDI Dopplergrams. This is important because these RVs are not models based on specific prescription. They are however due to active regions only and do not include all surface effects. Finally, a similar analysis on the RV reconstructed from HMI Dopplergrams provided by \cite{milbourne19} do not exhibit the residuals observed on HARPS-N RVs either.

\subsection{Search for an instrumental origin}

\begin{figure}
\includegraphics{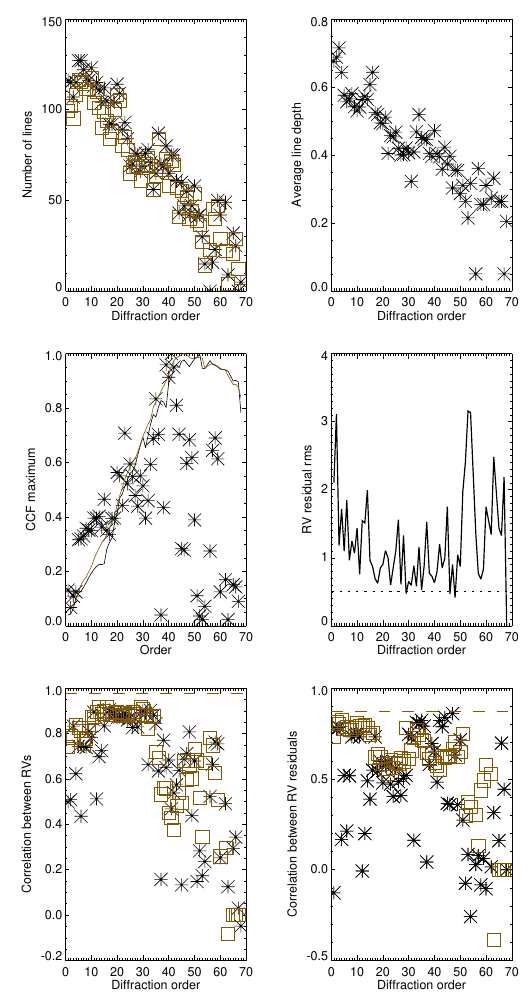}
\caption{Properties of the spectra, RV, and residuals versus {diffraction} order. Figure shows the number of lines in the G2 DRS mask in black and in the line-by-line analysis in brown (first panel), average line depth (second panel), CCF maximum (normalised to 1, stars) and flux in the spectrum (normalised to 1, solid line), rms of the residuals after subtraction of the S-index model (solid line) and same residuals on the full DRS time series (horizontal dashed line), correlation between RV for each {diffraction} order with the DRS RV based on CCF (black) and line-by-line analysis (brown), correlation between residuals (based on S-index model) for each {diffraction} order and the residuals for the full DRS time series based on CCF (black) and line-by-line analysis (brown). The horizontal brown dashed lines in the lower panels are the correlation between the reconstructed RV (respectively the residuals) on all {diffraction} orders for the line-by-line analysis and the DRS one. 
}
\label{ordre}
\end{figure}

We could not identify any solar effect which directly or indirectly (because the Sun is not point-like or because the Earth is orbiting the Sun) could explain these RV residuals.  Details are given in Appendix~\ref{append_solar}. In this section, we therefore explore the possibility for an instrumental origin of this effect. We also analyse the individual CCFs (one per {diffraction} order) and compute line-by-line RVs in order to get a better diagnosis: This allows us to evaluate the dependence of these RV residuals on {diffraction} order, wavelength and line depth and to identify other possible instrumental systematics. 

\subsubsection{Detector warm-ups}
\label{531}

The HARPS-N detector undergoes periodic warm-up to compensate for a small leak in the cryostat, leading to a progressive building-up of humidity \cite[][]{dumusque21}: this leak is responsible for  ghosts of increasing amplitude in the raw spectra, which impact the S-index measurements, hence the regular warm-ups performed to remove humidity. The detector temperature described by the keyword "HIERARCH TNG INS DETHDBODY\_T MEAN" 
is shown in the upper panels of Fig.~\ref{temp}, with the time of warm-ups indicated by the vertical dashed lines, while the middle panels show the RV residuals (after subtraction of the linear S-index model) for comparison.
The increase in RV is not as sharp on the smoothed residuals due to the smoothing but is visible on the time series with all points. We note a strong correspondence between the time of the warm-ups and the jumps in the RV residuals, as shown in the upper panel in Fig.~\ref{rvca}. The shape of the RV residuals also suggests a sharp rise followed by a slower descent, which would be compatible with this type of origin. When considering  observations before day 8100 only, the Pearson correlation between the temperature and the residuals is 0.49 on the smoothed series and 0.29 on the daily measurements. If we use this linear correlation to apply a crude correction to the residuals, it allows us to decrease slightly their rms, from 0.49 to 0.42 m/s for the smoothed residuals (Fig.~\ref{temp}). These new residuals are shown in the lower panels, where the jumps at the time of warm-ups are less pronounced. The impact is better seen on the periodograms, shown in Fig.~\ref{periodotemp}. In particular, after this correction, the power of the peak around 200 days is lower by about a factor of 2.5-3. It is therefore very likely that such warm-ups are responsible for important systematics, with an amplitude may be as large as 1.5 m/s peak-to-peak and affecting both RVs and BIS.

\subsubsection{CCF analysis and Line-by-line analysis: general overview}

In a first approach to study the wavelength dependence of this effect, we computed RVs from individual CCFs for all observations. The details of the protocol and some systematic effects are described  in Appendix~\ref{app_ccf}. 
In a second approach, we computed line-by-line RVs, which allow us to access more properties. The details are given in Appendix~\ref{app_lbl}.
Figure~\ref{ordre} shows the number of spectral lines, the average line depth and the amplitude of the CCF and flux versus {diffraction} order compared to the flux in the spectra (three first plots). 
The correlation between the order-by-order RVs, either based on the CCF of line-by-line analysis (lower left panel) with the global RVs is maximum for {diffraction} orders around 20-25, range which represents the combined effects of line depths, number of lines, and flux. However, this correlation drops towards the blue and even more towards the red: RVs derived from {diffraction} orders with a low correlation are therefore very different from the global RV time series.
Figure~\ref{series_ccf} illustrates the presence of strong systematic effects on each {diffraction} order, which cannot be explained by the difference in photon noise or line content only, and which are also affecting the long-term variability: The RV slopes versus time (see Appendices~\ref{app_ccf} and \ref{app_lbl} for details) show a systematic effect as a function of wavelength, in correlation with the variability of the continuum. In both analyses, we do not observe any systematic impact of {diffraction} order, wavelength or line depth on the signal at 200 days.

\section{Conclusions}

In this work, we performed a precise comparison between the model describing the contribution of active regions to RVs we used in \cite{meunier10a} with the solar HARPS-N RV time series \cite[][]{dumusque21} on short and long timescales. We also refined our diagnosis based on convective blueshift computation as well as the RVs in individual {diffraction} orders and lines. 
Our conclusions are two-fold. 

Firstly, we find that the prescription of 190 m/s for the convective blueshift used in \cite{meunier10a} should be multiplied by almost a factor of 2 to explain the variability seen in the solar HARPS-N RV time series. We expect such a prescription to be sensitive to the method used to measure RV (here the CCF, based on a certain list of lines) and to the definition of plages. This significantly  impacts the mass uncertainties and detection rates obtained in the blind tests performed in \cite{meunier23}, leading to a poorer performance. We also propose a simple prescription for the dependence on size of the convective blueshift in plages. However, there is a large dispersion among plages, associated with different responses to the chromospheric emission and to average magnetic field, which should impact the mitigating technique.  Finally, we also attempted to use these data to evaluate the attenuation factor of the convective blueshift in plages. This proved to be difficult to constrain, but points toward a factor of close to 1.

Secondly, despite the quality (and in particular a very low photon noise) of such solar observations, which is critical for benchmarking different approaches to deal with stellar activity, we found several significant systematic effects. 
The main one is the presence of a large amplitude signal (about 2 m/s peak-to-peak), already seen in \cite{dumusque21}. We checked many possible solar sources, none of which could explain a signal of this magnitude. We found that a significant part, if not all, of this signal is likely to be due to the periodic warm-ups of the detector. We have been able to characterise it, showing that it was also strongly seen in the BIS. It affects lines of all depths and at all wavelengths. The artefacts created by these warm-up should significantly impact correlation with various activity indicators \cite[such as those studied in][]{sen23}. 

In addition, different {diffraction}  orders  present very different long-term trends (including a change in sign) that are correlated with the continuum variability. This casts some doubt on the exact amplitude of the long-term trend in global RV. In principle, this should also impact the methods used to mitigate stellar activity based on subsets of lines. Finally, we also suggest that an offset of about 1.4 m/s between the time series acquired before the interruption and after is very likely. Therefore, instrumental systematics still pose a significant limitation for high-precision studies. A comparison between time series obtained with different instruments, presented in \cite{zhao23b}, open up a novel way to combine them. In this context, it could prove extremely interesting to identify and mitigate those systematic effects in a future study.

\begin{acknowledgements}

We thank Christophe Lovis for providing the ESPRESSO G2 masks. 
This work was supported by the "Programme National de Physique Stellaire" (PNPS) of CNRS/INSU co-funded by CEA and CNES.
This work was supported by the Programme National de Plan\'etologie (PNP) of CNRS/INSU, co-funded by CNES. We made use of the BASS2000 website. HMI and MDI data were retrieved from the JSOC archive. We thank Arthur Amezcua for his help in retrieving the data.  data for the calibration of the spots extracted for HMI maps have been provided by USAF/NOAA.
SOHO is a mission of international cooperation between the European
Space Agency (ESA) and NASA. We used SILSO data/image, Royal Observatory of Belgium, Brussels. The HARPS-N solar data were retrieved from the DACE platform. Ca K index was
provided by the Sacramento Peak Observatory of the US Air Force Phillips
Laboratory (http://nsosp.nso.edu/cak\_mon). This work has made use of the VALD database, operated at Uppsala University, the Institute of Astronomy RAS in Moscow, and the University of Vienna. We thank for exchanges on the convective blueshift measurement and Jesper Schou for information on MDI RV calibrations.  
\end{acknowledgements}

\bibliographystyle{aa}
\bibliography{biblio}

\begin{appendix}

\section{Comparison between the original cycle 23 models and cycle 24 observations}
\label{cyc23_24}

\begin{figure}
\includegraphics{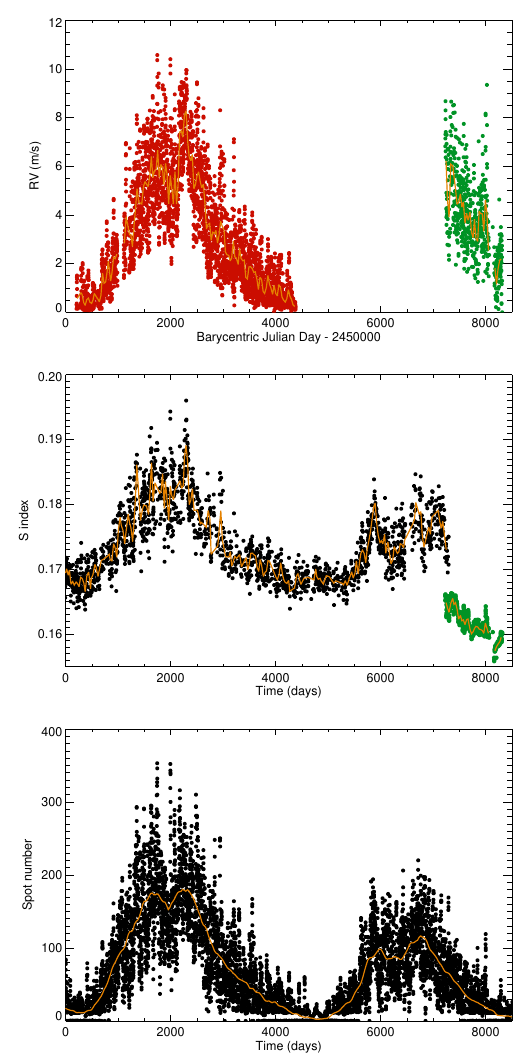}
\caption{ 
Time series comparing cycle 23 model and cycle 24 observations. The orange lines represent smooth time series from a running average (over 28 d). The upper panel shows the model RV from \cite{meunier10a} in red and the HARPS-N observed RV in green. The middle panel shows the Sac Peak S-index (black) and the HARPS-N S-index (green). The last panel shows the Sunspot number from SILSO/SIDC (https://www.sidc.be/silso/) over cycles 23 and 24 . }
\label{serie}
\end{figure}

Figure~\ref{serie} illustrates the comparison between our 2010 model obtained for solar cycle 23 and the HARPS-N observation over three years, corresponding to the descending phase of cycle 24. A simple visual examination shows that the RV trend observed on the HARPS-N RVs is slightly stronger than the slope observed during the descending phase of cycle 23 in our model, despite cycle 23 being significantly more active (as shown in the lower panel). This qualitative comparison suggests that the reconstruction performed in 2010 underestimated the amplitude of the RV signal due to the inhibition of the convective blueshift. Comparison of the two RV time series based on their respective relation with the S-index is challenging however. First, the Sacramento Peak data, which can be used to analyse the cycle 23 models, are much noisier than the HARPS-N data. In addition, there is an offset and possibly a scaling factor between the Sacramento Peak  and the HARPS-N indices, but the overlap between the two time series is unfortunately too poor to determine if there is a scaling factor or not.

\section{Models and calibrations}

\subsection{Calibration for spot and plage extraction from HMI/SDO images}
\label{app_calib}

HMI magnetogram and intensity maps (the latter corrected from limb darkening) were retrieved from the JSOC data base with the {\it drms} package \cite[][]{glogowski19}. In order to ensure that the sizes of the spots and plages extracted from those maps correspond to the measurements applied in 2010, we calibrated the threshold  to be applied to  HMI map as follows.

We used HMI intensity maps between July 2015 and June 2017 for the days with HARPS-N observations, and computed the spot filling factor based on different intensity threshold between 0.7 and 0.9 (the quiet Sun corresponds to 1). We then compared it with the filling factor of spots extracted from the USAF catalogue and chose the best intensity threshold, 0.80.

Concerning plages, in 2010, we used MDI magnetograms and a threshold of 100~G, providing a list of structures corresponding to plages and magnetic network. HMI magnetograms are different and therefore we need to identify the threshold that would provide an equivalent filling factor of plages compared to MDI data. For that purpose, we retrieved 60 pairs of magnetograms (MDI and HMI) over one year, between April 2010 and April 2011, both maps being taken at exactly the same time. HMI magnetograms were binned to the MDI resolution (from 4096x4096 maps to 1024x1024 maps). We tested  different thresholds applied to the HMI magnetograms between 40 and 100~G,  leading to a threshold of 5.7~G providing the best agreement between the plage filling factors found for both instruments.

\subsection{Cycle 24 spot and plage catalogue}
\label{app_cyc24}

We retrieved HMI magnetogram and limb-darkening corrected intensity maps each day with HARPS-N observations. They were chosen to be simultaneous, and as close as possible to the average time of HARPS-N observations. 
There is one day of HARPS-N observations with no HMI reliable data, so that the analysis is made over 602 days. 
Following the calibration described in the previous section, we applied the same protocol to provide a list of spots (position on the disk and size in ppm of the solar hemisphere) for each day. 

Magnetograms were binned to match MDI resolution, and a threshold of 55.7~G was applied on the absolute value of the magnetic field. As in 2010, structures smaller than 4 pixels were removed to avoid being impacted by noise. However, we also removed the spots in order to retrieve plages only. This also provides a list of plages, including network features (position on the disk and size in ppm of the solar hemisphere). In addition, the average of the absolute value of the magnetic field, |B|$_{\rm disk}$, is computed over the whole disk.

\subsection{Velocity dependence on plage size}
\label{app_depa}

\begin{figure}
\includegraphics{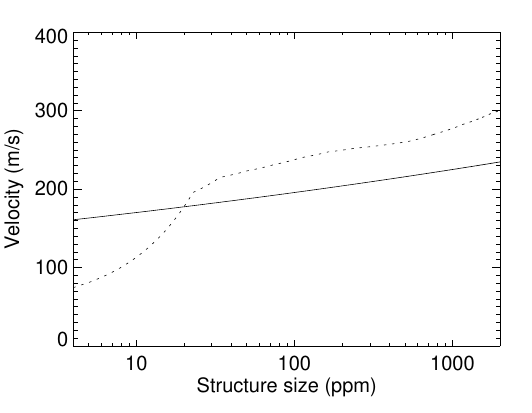}
\caption{
Velocity versus size (in ppm of the solar hemisphere) used in \cite{meunier19} (dotted line) and after the new analysis (Eq.~\ref{eqpowerlaw}). Both were normalised so  that when applied to the size distribution of the structures (corresponding to HARPS-N observations) in this paper, they are equivalent to 190 m/s.
}
\label{law}
\end{figure}

In the simulations performed in \cite{meunier19}, we used a size-dependence of the velocity plages based on the results obtained in \cite{meunier10}.
The curve was different in practice from  the one shown in \cite{meunier10} due to a shift in size:
The velocity was dropping for structures smaller than 20 ppm of the solar hemisphere while the threshold should have been lower: It is likely that in this case the contribution of the smallest structures was slightly underestimated. 
We also used the RVs per structure extracted from MDI Dopplergrams \cite[][]{meunier10} and applied a Gaussian fit on the distribution of values in each size bin for a better robustness to derive a new law. Both laws were then scaled so that given the typical distribution of structure sizes in these simulations, the average was equivalent to the original prescription of 190 m/s. The resulting velocity versus size was fitted with a power law. Both laws are shown in  Fig.~\ref{law}.

\section{Attenuation factor in plages}
\label{fact_att}

A first possibility to evaluate the attenuation factor of the convective blueshift is to use the slope of $\gamma$ versus the filling factor (ff) and extrapolate to ff=1. We  describe the (global) convective blueshift ($CB$) with two components, a quiet one and an active one: 

\begin{equation}
CB=CB_0 \times (1-{\rm ff}(t))+A \times CB_0 \times {\rm ff}(t),     
\end{equation}
where $CB_0$  is the quiet Sun value. $A\times CB_0$ represents the convective blueshift in plages and therefore 1-$A$ is the attenuation factor. Based on the work of \cite{gray09} suggesting that the third signature is universal, we propose a similar equation by replacing $CB$ by $\gamma$ ($\gamma$=0 would correspond to a configuration with no convective blueshift), so that $\gamma$ provides direct information on the convective blueshift:

\begin{equation}
\gamma=\gamma_0 \times (1-{\rm ff}(t))+A \times \gamma_0 \times {\rm ff}(t).     
\end{equation}

In this two-component model, we assume that ff includes all structures impacting the CB, and these are subject to the same attenuation factor. The  attenuation factor is then equal to the opposite of the slope divided by $\gamma_0$. 
We performed a linear fit on $\gamma$ versus ff for a 55~G threshold (Fig.~\ref{tss2}), which gives an attenuation factor of 1.25$\pm$0.23. This value is larger than 1, which is not physically possible, although it could be marginally compatible with 1. 
This is probably due to the fact that this definition of ff may exclude a fraction of the surface also impacting the CB, but corresponding to lower magnetic fields. A lower threshold of 23~G (which corresponds to a threshold of 40~G on MDI magnetograms) leads to the curve in red, and an attenuation factor of 0.58$\pm$0.09. The median ratio between the filling factor defined by the 55~G (black curve in upper panel of Fig.~\ref{tss2}) and 23~G (red curve) threshold is around 3.
The use of a lower threshold to define the ff (here 23 G) ensures that the assumption that the surface outside the considered structures exhibit convective blueshift which is not attenuated is good, however the added surface probably includes regions where the attenuation factor should be much lower than in large plages. The resulting attenuation could therefore represent a lower limit for large plages.

An alternative solution relies on the prescription obtained in Sect.~3 for the amplitude of the inhibition of the CB in plages. A prescription of 359 m/s (prescription used in 2010 time the correcting factor of 1.89) applied vertically in each point of the surface as in our model  corresponds to 254 m/s when integrated over the solar surface due to projection effects. If we know $CB_0$ corresponding to the quiet Sun, then the attenuation factor is 254/$CB_0$. 
We use the $CB_0$ of 267 m/s derived in Sect.~3.3 from our velocity versus $d$ law and a weighting in $d$ of the G2 mask lines to compute the weighted-average $CB_0$. This gives an attenuation factor of 0.96, namely, a value that is very close to 1. However, any error on the zero would impact this ratio, as discussed in Sect.~3.3. The uncertainty on the proper choice of weighting also impacts this estimation. We note however that the weighting in $d^2$ leads to a $CB_0$ lower than the prescription, which is not physically possible either, so that this weighting may not be adequate (if the zero is correct).  This estimation therefore depends on the appropriate weighting. 
Applying this average factor to the dependence in Eq.~\ref{eqpowerlaw} leads to a attenuation factor of 1.1 (again higher than 1, so not plausible, but pointing towards a strong attenuation factor) for the largest structures and 0.73  for the smallest ones.

\section{Search for a possible solar origin of the peak at $\sim$200 days}
\label{append_solar}

\begin{figure}
\includegraphics{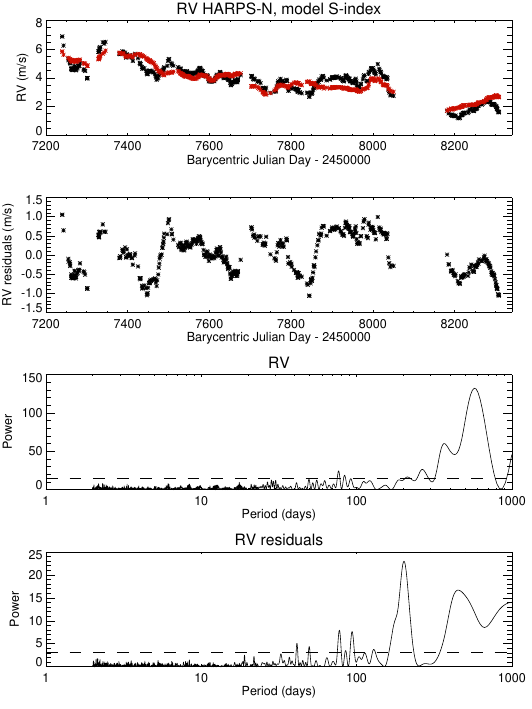}
\caption{
Smoothed HARPS-N RV (black), superposed to the model based on the S-index (red), the residuals (difference between the two), periodogram of the HARPS-N RV, and periodogram of the smoothed residuals.
}
\label{pic200obs}
\end{figure}

We  explored several possibilities based on solar phenomena to attempt to explain the periodicity around 200~d observed in the RV residuals after subtraction of our model (see Sect.~\ref{sec3}) or a model based on the S-index (see Sect.~\ref{sec51}). 

\paragraph{Solar phenomena at this typical periodicity}
The dominating cycle is by far the Schwabe cycle of 11~yrs, which corresponds to magnetic field reversals every 22~yrs. The spot number also exhibits a much longer periodicity of about ~90~yrs corresponding to the Gleissberg cycle \cite[][]{gleissberg39,gleissberg55}, which corresponds to a modulation of the amplitude of the Schwabe cycles. 
On the other hand, variability at shorter time-scales has also been observed, with Rieger-type periodicities of a few months around 150~days seen in various indices \cite[e.g.][]{rieger84,lean89}, mostly based on flares and a few proxies (but not with plage proxies), or with a quasi-biennal oscillations \cite[e.g.][]{sakurai81,vecchio09}, first detected based on neutrinos and then with coronal lines. None of those phenomena are strictly periodic, but are rather the superposition of stochastic variability. Despite the abundant literature on solar variability based on a very large number of processes and indicators, there is however no indication of 180 or 200 d periods. In addition, periodogram of the filling factor, |B|$_{\rm disk}$, or spot number over the same period than HARPS-N solar observations do not show this periodicity. 

\paragraph{Impact of $B_0$ on the active region contribution. } 
Because the signal is not seen in the MDI and HMI Dopplergram reconstructed RVs, a contribution of active regions (either through contrast or convective blueshift inhibition) is unlikely. However, given the periodicity close to 180 d, we quantified the effect the varying $B_0$ over $\pm$7.25$^\circ$ over the year by computing our models with this variation, and compared them with the edge-on models. 
The rms of the differences between RV time series is  0.12 m/s for cycle 23 and 0.04 m/s for cycle 24 only, so that this cannot be the origin of the RV residuals.

\paragraph{Impact of the solar $B_0$ angle on the meridional circulation contribution}
Similarly, we checked the impact of a varying $B_0$ solar angle over time on the integrated meridional circulation. We used the average latitudinal profiles derived from the observations by \cite{ulrich10} studied in \cite{meunier20c}.
We observe a clear variability with a 180 d period, however a small amplitude of only 0.23 m/s peak-to-peak, which is insufficient to explain the observations. 

\paragraph{Impact of other known solar processes}
The other global process affecting RVs is supergranulation, but with an amplitude below 1~m/s and very weak  latitudinal dependence, it cannot be responsible for the observed residuals. This is even more true for granulation and oscillations. 

\paragraph{Impact of airmass}
Because the Sun is not point-like, the inclination of the solar rotation axis with respect to the vertical in the sky combined with solar rotation leads to some spurious RV residuals during the day. This effect has been taken into account in the RVs computation \cite[][]{collier19}. We therefore do not expect strong effects due to airmass, but since rotation is the main remaining reservoir in terms of solar velocity field, we checked if airmass could impact these residuals, possibly due to some other effects (in particular because there is no atmospheric dispersion compensator). The elimination of the points with the largest airmass does not change significantly the amplitude of the residuals, and the periodicity of the  residuals is 1 year and not 200 days. In addition, we plotted the BIS as a function of airmass, and found a dependence with a small amplitude (below 0.2 m/s), which is much smaller than the observed effect. 

We conclude that this variability does not have a solar origin. However, solar variability could contribute for a small fraction of those residuals.

\section{Detector temperatures}
\label{app_temp}

\begin{figure}[h]
\includegraphics{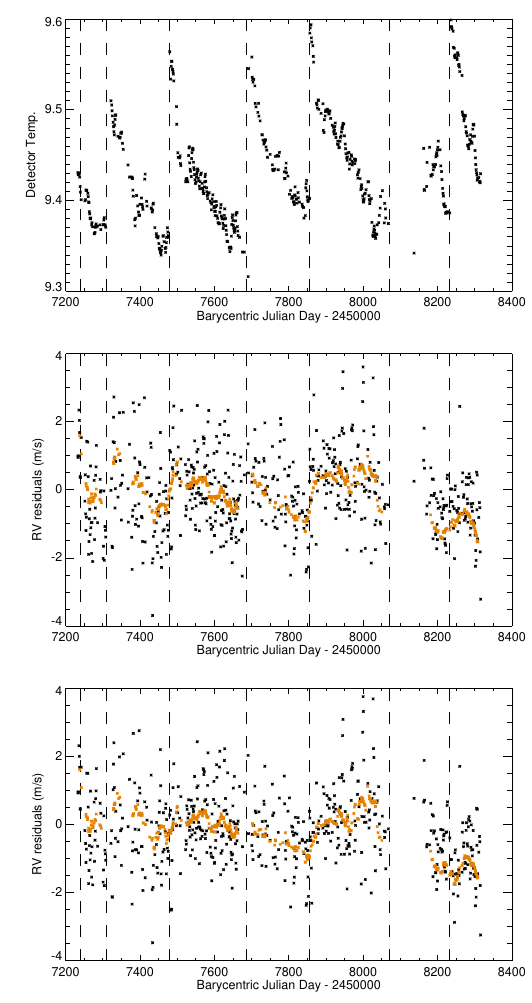}
\caption{Daily time series of the detector temperature from the file headers (upper panel), of the RV residuals based on a linear model based on the S-index (middle panels) and after an additional correction based on the detector temperature (lower panels). The orange dots correspond to the smoothed residuals (over 28 days). 
}
\label{temp}
\end{figure}

\begin{figure}
\includegraphics{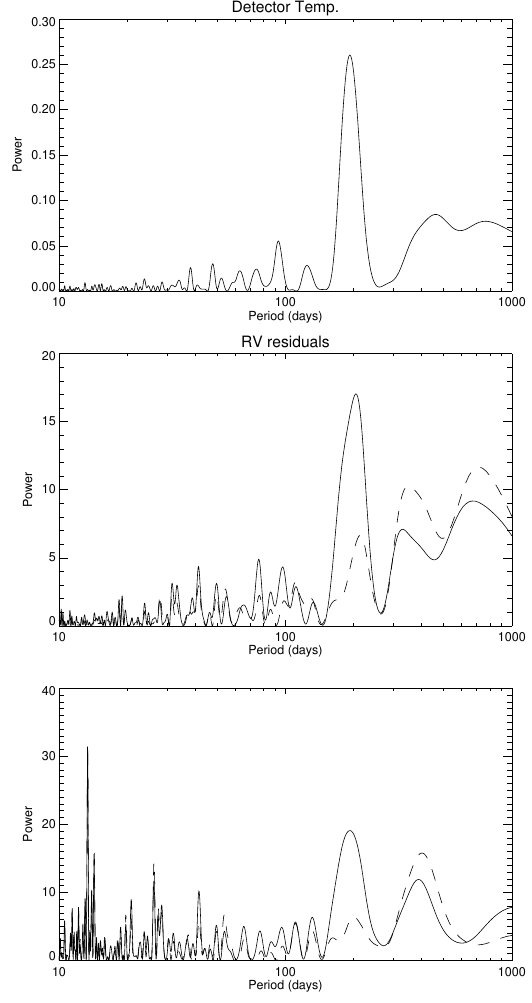}
\caption{Periodograms of the detector temperature (upper panel) and of the daily RV residuals (middle panel) and smoothed time series over 28 days (lower panel):  after S-index model subtraction (solid line), and after an additional correction based on the detector temperature (dashed line). 
}
\label{periodotemp}
\end{figure}

We compare the RV residuals and the detector temperature in Fig.~\ref{temp}. The periodograms are shown in Fig.~\ref{periodotemp}. These results are discussed in Sect.~\ref{531}).

\section{Complementary  RV computations}
\label{app_calcrv}

We detail here the method used to analyse the spectra. These analysis allows us to  discuss several instrumental systematics. 

\subsection{RV with respect to laboratory wavelengths}
\label{app_tss}

\begin{table*}
\caption{Laboratory wavelengths (extract)}
\label{lambda}
\begin{center}
\begin{tabular}{llllll}
\hline
Line & $\lambda$ & Reference & $\lambda$ & Reference & Line depth \\
  &  (2017+) & & (VALD) & & \\
  \hline
  FeII&4024.5502& (2)&4024.5505& (8)& Not meas.\\
FeII&4178.8547& (2)&4178.8536& (8)& 0.74\\
FeII&4384.3138& (2)&4384.3129& (8)& Not meas.\\
FeII&4413.5920& (2)&4413.5912& (8)& 0.36\\
FeII&4416.8196& (2)&4416.8186& (8)& 0.68\\
\hline
\end{tabular}
\end{center}
\tablefoot{Line used in the convective blueshift analysis (Sect.~\ref{app_tss}). The first set of lines (column 2) were those used in \cite{meunier17,meunier17b}, complemented by Ni lines in this work. The references for the wavelengths are: (1) \cite{dravins08}, (2) \cite{nave94}, (3) \cite{litzen93}. The second set of wavelengths were retrieved from VALD and the corresponding references are: (4) \cite{K14}, (5) \cite{BK94}, (6) \cite{FMW}, (7) \cite{BWL}, (8) \cite{K13}, (9) \cite{BSS80}, (10) \cite{K08}, (11) \cite{WLSCow}, (12) \cite{K16}, (13) \cite{KL}. The full list is available at the CDS.
}
\end{table*}

Based on S1D spectra, we computed the RVs with respect to laboratory wavelengths in order to characterise the convective blueshift based on the position of the center of spectral lines \cite[][]{gray09,reiners16,meunier17,meunier17b,liebing21,almoulla22} to compare the expected convective blueshift of the MDI line with respect to other lines and to directly evaluate the variability over time of this convective blueshift. 
We used a list of spectral lines of FeI lines \cite[][]{nave94}, TiI and FeII lines \cite[][]{dravins08}, and  as in \cite{meunier17,meunier17b}, to which we also added Ni lines from \cite{litzen93}.
This represents an initial list of 2518 spectral lines, 80\% of which are FeI lines. The wavelengths are listed in Table~\ref{lambda}. The position of the lines in each S1D spectra is determined by a polynomial fit around line center, following \cite{gray09} in order to reproduce the third signature. The resulting velocity for each line at each time step was then computed, and we subtracted the variability based on the raw velocity provided by the DRS (with no global offset). 
We also checked the result on the same set of lines but with laboratory wavelength from VALD, also listed in Table~\ref{lambda}. The results were similar, with a global curve slightly higher and a sightly strong curvature (shift of 25 m/s for $d$=0 and 39~m/s for $d$=1).

We analysed the velocity as a function of line depth (also determined from the spectra), and removed the outliers a posteriori as follows. 
We first eliminated the strongest outliers based on a 5-sigma clipping approach applied to each line \footnote{The threshold eliminates strong outliers only, but the exact choice is not critical because the subsequent analysis is performed on distributions of RVs}. We then computed the time-average line depth $d$ and velocity for each spectral line, and applied a correction of the wavelength dependence from \cite{liebing21}. This correction was used only to make the selection of outliers more robust but was not removed from the final velocities. The distribution of those averaged velocities in each bin in $d$ was fitted with a Gaussian, to determine a 3-$\sigma$ level based on the fitted width: Lines outside this range were then eliminated from the analysis. Because very weak lines present a huge dispersion and large uncertainties, we consider lines with $d$>0.1 only. This leads to 1341 usable lines. 
After this line selection, we also removed outliers corresponding to individual measurements: For each set of values for a given line, we also applied the threshold at the 3-$\sigma$ level, using the width found above for the corresponding $d$ bin.
To study the daily values, we then selected valid measurements for this day, binned them (with a bin of 0.1 in $d$), performed a Gaussian fit on their distribution to find the best velocity for each $d$ bin. Those RVs are then analysed to find $\gamma$ at each time step (see Sect.~\ref{sec33}), namely the ratio to be applied to a function in $d^3$ compared to the fit performed by \cite{liebing21}. A large value of $\gamma$ means a strong dependency of the convective blueshift on $d$.

\subsection{CCF RVs per {diffraction} order}
\label{app_ccf}

\begin{figure}
\includegraphics{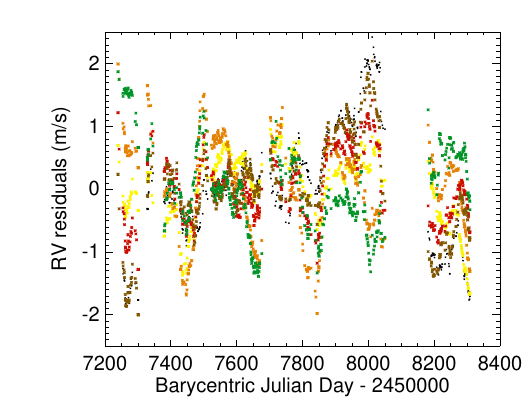}
\caption{Smoothed residuals after the subtraction of the S-index model versus time for six ranges of {diffraction} orders, based on the CCF analysis: orders 1-11 (black), orders 12-22 (yellow), orders 23-33 (orange), orders 34-44 (red), orders 45-55 (brown), and orders 56-69 (green). 
}
\label{ordrepaq}
\end{figure}

\begin{figure*}
\includegraphics{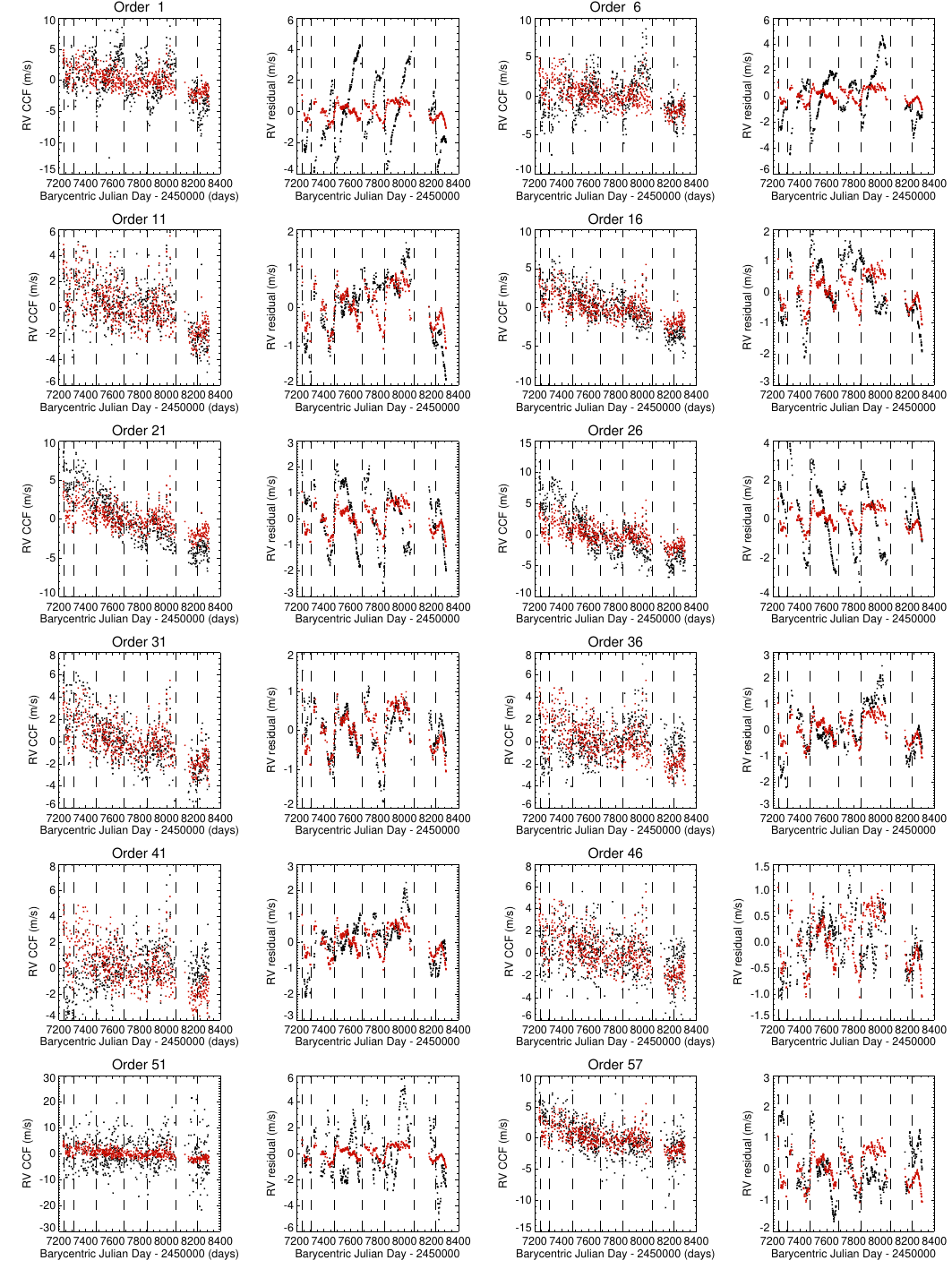}
\caption{Daily time series of the CCF RV and smoothed RV residuals (after subtraction of the S-index model) versus time for a selection of {diffraction} orders across the spectrum (in black). The red dots correspond to the global time series. The vertical dashed lines indicate times of detector warm-ups. 
}
\label{series_ccf}
\end{figure*}

\begin{figure*}
\includegraphics{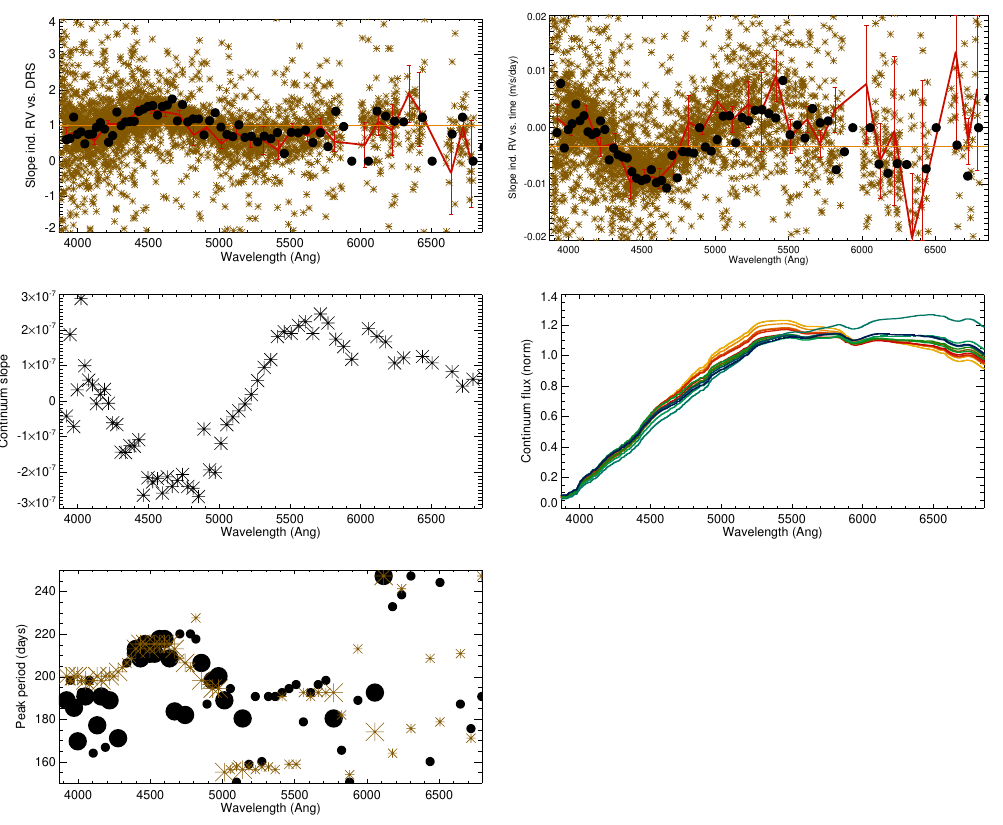}
\caption{Slope of the individual RVs versus the DRS RV (upper left panel) and time (upper right panel) as a function of wavelength: CCF value for each {diffraction} order (black, error bars are plotted but smaller than the symbol size), individual line-by-line values (brown), binned values on line-by-line values (red), value for the DRS time series (orange horizontal lines). The ranges in ordinate have been restricted (outliers lie outside the visualised range, 9\% in the upper panel and 16\% in the lower panel). The slopes of S$_{\rm cont}$ versus time are shown in the second row (left panel). The right panel in the second row shows 13 examples of daily continuum spread over time (from the beginning of the observations in yellow to the end in blue). The last panel shows the peak period (selected in the 150-250 days, the largest symbols corresponding to {diffraction} orders for which the peak in that period range is the highest peak in the periodogram) versus wavelength for the different {diffraction} orders: CCF analysis (black) and line-by-line analysis (brown). 
}
\label{penteslam}
\end{figure*}

We applied a Gaussian fit on each individual CCF retrieved from the DACE archive (one per {diffraction} order at each time step), from which we derived daily time series. 
We checked that when averaging them with a weight corresponding to the amplitude of each CCF, the result was very similar to the RV computed on the global CCF (that is after summing all {diffraction} orders) with the DRS: The rms of the difference is 0.06~m/s only, which is small compared to the RV residuals we wish to study.

The RV time series for each {diffraction} order were then binned (with a running mean) over 28 days to compute the residuals as for the global time series in order to study the behaviour of the peak at $\sim$200 days, as in Sect.~\ref{sec51}. The linear fit of each RV time series as a function of the S-index provide a model which is subtracted, allowing us to compute this RV residuals separately  for each {diffraction} order. We then computed the amplitude of those residuals and their periodogram.  We also regrouped the time series into 6 groups of {diffraction} orders, from the blue to the red, each group containing 11 (with 10 for the last bin) orders. They were then analysed with the same procedure. 

Figure~\ref{ordrepaq} shows the residuals (after subtraction of the S-index model and smoothed, see Sect.~5.1) for the 6 ranges of {diffraction} orders in black compared to the residuals for all orders (based on the DRS RV) in red: There are strong similarities, with no obvious effect as a function of wavelength. 

We now discuss  a few properties of these individual RVs. Some {diffraction} orders, mainly on the red side of the spectrum, behaves in a manner that is very different from the other orders, with very low correlation of the RV time series with the DRS ones (as summarised in Fig.~\ref{ordre}). Examples of time series individually for several {diffraction} orders  are shown in Fig.~\ref{series_ccf}. 
We computed the RVs based on a selection of {diffraction} orders with a threshold defined by the correlation between RVs:  considering orders with the highest correlation and then including orders with progressively lower correlations, we find that  adding orders with correlations below 0.4 (about 10 orders) does not add any new information and does not contribute to decrease the uncertainty on the RVs. Improvements for those orders could however come from a correction of tellurics in those orders, for example such as proposed by \cite{ivanova23}.

We also compared the RV computed for each {diffraction} order with the DRS RV time series by computing the slope between the two time series. A good agreement corresponds to a slope of 1. The results are shown in the upper left panel of Fig.~\ref{penteslam} in black. We find a systematic variation of that slope with wavelength, with a stronger variability around 4600 \AA, and a low variability in the bluest {diffraction} order and around 5200 \AA. Another way to observe this is to compute the slope of RV versus time (since the data cover only 3 years and are dominated by a trend). This slope, shown in the upper right panel of Fig.~\ref{penteslam}) is negative for the DRS RV. This is interesting because this slope may give information on additional process (Sect.~\ref{sec34}).  We find a similar variability with wavelength (naturally anticorrelated with the curve in the upper left panel), including {diffraction} orders showing a reversal of the long-term trend.

We investigated this effect further to understand the origin of this behaviour, by analysing the shape of the continuum of the spectra over time, as such changes can affect the computation of the line positions. For each spectrum, we computed the slope S$_{\rm cont}$ of the flux in the continuum versus wavelength. S$_{\rm cont}$ was then normalised  by the flux in each {diffraction} order to be able to compare spectra with different fluxes over time. 
These slopes were then binned over each day, and the slope of S$_{\rm cont}$ versus time was computed: This slope is shown in the third panel in Fig.~\ref{penteslam} versus {diffraction} order and exhibit a dependence on order which is similar to the  first panels. The fourth panel illustrates the continuum for a few examples spread over the duration of the observations:
The flux level as a function of wavelength, commonly known as the spectral colour, significantly changes as a function of time. Such a change could possibly be due to variability in stray light over the detector. The ESPRESSO DRS corrects for background contamination (stray light) at the level of the raw images before extraction, however, in this process, ghosts, that are known to significantly vary on both sides of a warm-up, are not considered.  At first order, we do not expect such a colour change to affect the final RVs, as the ESPRESSO DRS corrects for any change in colour of the extracted spectrum, by rescaling each {diffraction} order with respect to a static template before computing the RV through a cross-correlation function. 
However, colour is corrected order-by-order and not whiting each {diffraction} order. Colour change within orders could induce a different weighting of lines in the CCF and could introduce a RV offset. A deeper analysis is out of the scope of this paper, however we point out here a possible direction in which to investigate further.
S$_{\rm cont}$ presents significant long-term variability do not exhibit any periodicity around 200 d. The properties of the continuum therefore do not appear to be responsible for the RV residuals studied in Sect.~\ref{sec5}. However, they may impact some of its properties. The peak at $\sim$200 days, based on  the Lomb-Scargle periodograms computed on the smoothed residuals, indeed exhibit different properties for different {diffraction} orders. The amplitude covers a large range with no systematic trend, but the period, despite some dispersion, exhibits a variability (lower panel in Fig.~\ref{penteslam}) below 5700~$\AA,$ which presents a similarity with the systematic effect with wavelength observed on RV variability and continuum properties.

We conclude that the time series for different {diffraction} orders appear to be affected by strong systematic effects with variability that cannot be due to a degraded photon noise when considering individual CCFs.  This is also seen with the large rms on the RV residuals, and the very low correlations for certain {diffraction} orders in the lower panels of Fig.~\ref{ordre}.

\subsection{Line-by-line RVs for the G2 mask}
\label{app_lbl}

\begin{figure}
\includegraphics{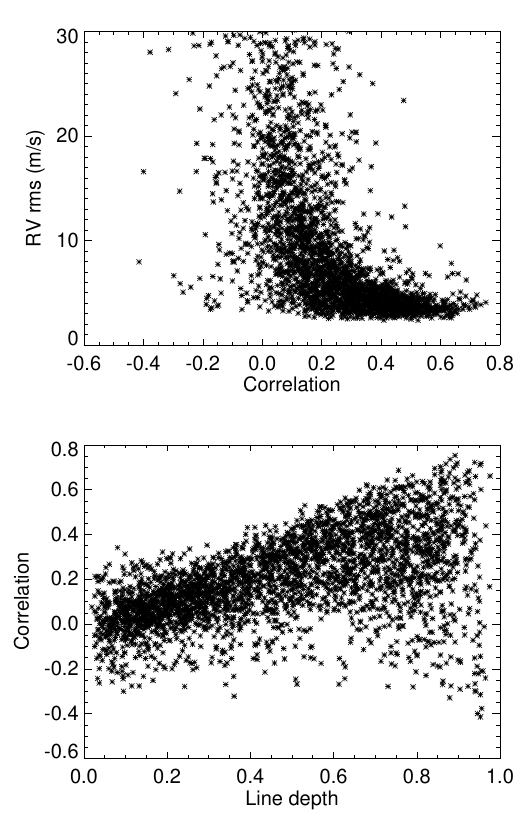}
\caption{Rms RV of the line-by-line RV time series versus their correlation with the DRS global RV (upper panel) and correlation versus line depth (lower panel). 
}
\label{proplbl}
\end{figure}

\begin{figure}[h]
\includegraphics{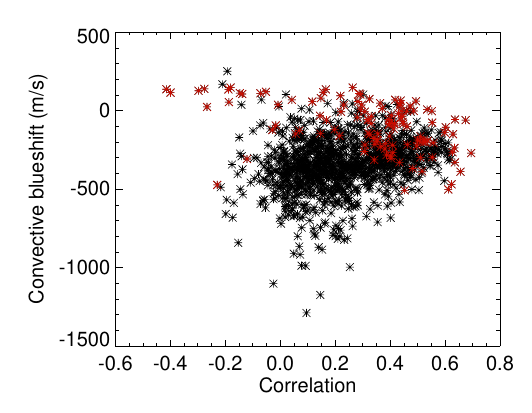}
\caption{
Convective blueshift (Sect.~\ref{app_tss}) versus correlation between line-by-line RV and DRS RV (Sect.~\ref{app_lbl}) for the 1224 lines in common. The points in red correspond to the 119 lines with $d$>0.8.
}
\label{testred}
\end{figure}

An average spectrum was computed using all spectra and used as a reference. It was also used to provide a preliminary list of lines (7357 lines before selection). Each S1D spectrum was then analysed as follows. The continuum was computed as in \cite{meunier17} and  subtracted. The RV for each line was computed following the procedure in \cite{bouchy01}, \cite{dumusque18}, and \cite{artigau22}, namely, by approximating the wavelength shift as the derivative of the reference spectrum times the step in wavelength. This was converted into velocity for each pixel and weighted-averaged over each line. The  uncertainties on each pixel of the S1D spectrum were used to compute the uncertainty on each RV value. 
After elimination of outliers based on a 5-$\sigma$ clipping approach and selection of lines present in the G2 mask used in the DRS (leading to 3112 usable lines out of the 3625 lines in that mask), the RVs were corrected from the daily drift (assumed to be wavelength-independent) and from the Solar System planetary signal with the correction provided by the DRS \cite[][]{collier19,dumusque21}. The velocities for each line were then averaged over each day to produce a daily time series for each valid spectral line. 
Time series were then binned over 28 days (with a running mean) to compute the residuals after subtracting the model based on the S-index as in Sect.~\ref{sec51}. This was done either individually for each line, or after regrouping them by {diffraction}  order, line depth or wavelength. We also checked that when combining all lines, there was a good agreement with the DRS RV: we combined them by fitting  Gaussian on the distribution of RVs at each time step, to better eliminate remaining outliers, weighting them with $d$ or $d^2$. For example, with the weighting in $d^2$, the correlation between this time series and the DRS RV is 0.98.

We do not find any dependence on line depth of the RV residuals. Regrouping them in 0.1 bins in line depth for line depth between 0.1 and 0.9 shows very similar residuals. These sets of lines being independent from each other, it shows that there is an underlying signal present everywhere in the spectra. We also checked the residuals for lines in the middle of each {diffraction} order compared to lines on the edges, since those might be more susceptible to be sensitive to wavelength calibration errors: There is no impact on the RV residuals either. 
After regrouping those lines per {diffraction} order (or in 6 groups of orders as for the CCFs), we find a strong variability of the residuals, which is  similar to what was observed with the CCFs. The 6 residuals are illustrated in Fig.~\ref{ordrepaq},  although other effects are present, illustrating the differences between {diffraction} orders. As for the CCF analysis, Fig.~\ref{ordre} shows that the correlation between RV residuals (lower panels) and the global residuals may be dominated by other systematic effects: All lines exhibit the residuals, but other effects affecting the lines such as the systematics discussed below makes the analysis as a function of wavelength difficult. 

One objective of this line-by-line approach was to evaluate to which weighting the DRS RV were equivalent to. This is useful in Sect.~\ref{sec33} to combine the convective blueshifts for the different lines to compute an equivalent convective blueshift. However, the different weightings give similar RV time series, which should not be the case if their variability was strongly depending on line depth. Two RV time series, computed separately for lines with $d$>0.5 and $d$>0.5, are also  very similar. When considering that the convective blueshift is attenuated in plages however, it is usual to consider that weak lines will exhibit a larger variability because their convective blueshift is stronger (so that a given fraction of it will naturally  lead to a stronger variability compared to deep lines). Such an expected behaviour was at the origin of the method based on subsets of lines proposed in \cite{meunier17c} to mitigate the contribution of the convective blueshift. The search for lines that are more or less sensitive to active regions also motivated the works of \cite{dumusque18} and \cite{cretignier20} for example.  This is not observed here because the whole line is used to compute RVs: this is consistent with the results of \cite{gray09} showing that the third signature was seen when considering the bottom of the lines.  \cite{almoulla22} performed a line-by-line analysis of the same HARPS-N solar data based on a much stricter selection of lines  and importantly do not compute RVs on the whole line but in spectral regions corresponding to different temperatures. They found an increasing rms RV with increasing T$_{\rm eff}$ (without counting the lowest T$_{\rm eff}$, which behaved differently); however, the effect remains small and the bin with the largest rms RV include less lines, which are also weaker, so that we expect a larger amount of noise.

As for the CCF, we study here a few properties of the individual RVs. The slope derived from the comparison with the DRS RV and versus time shows the same wavelength-dependence than for the CCF (brown symbols in Fig.~\ref{penteslam}). The variation of the slope versus time is problematic because it impacts the definition of the trend in the final DRS RV, and therefore the robustness of the long-term time series. It is striking that the slope (versus time) is close to zero and even positive (conversely to the global DRS RV) for {diffraction} orders corresponding to the maximum of flux, where we would expect the most reliable RVs. 

Finally, Fig.~\ref{proplbl} shows the rms of the individual time series versus their correlation with the DRS RV and the relation between this correlation and line depth. 
Only   7\% of the lines exhibit a correlation higher than 0.5, and they correspond to the deepest lines: this is likely biased by the fact that the strongest lines are also the most precise ones. However, a large proportion of lines exhibits a low correlation, associated with the fact that many of them display a very large dispersion of RV values over time. 
In addition, we selected the 1224 lines that were also in the list of laboratory wavelengths (Sect.~\ref{app_tss}) and retrieved their convective blueshift, illustrated in Fig.~\ref{testred}. There a slight trend for the deepest lines ($d$>0.8), shown in red, to have an apparent convective redshift to present some anticorrelation. However, a visual examination of the corresponding time series shows that this does not correspond to a reversed variability compared to those with convective blueshift: It is in fact dominated by a reversed long-term trend, not necessarily due to the convective blueshift. This may also be due to the comparison with the reference spectra, as lines moving more than average could lead to an anticorrelated behaviour:
\cite{almoulla22} found a possible anticorrelation for lines with the lowest T$_{\rm eff}$ (corresponding to the deep lines here), including based on the long-term term. However, this T$_{\rm eff}$ bin included a lower number of lines than the other bins, and given the effects with wavelength mentioned above, we cannot exclude an artefact, depending on the wavelength of those lines. 

\end{appendix}

\end{document}